\documentclass[prd,12pt,nofootinbib,preprint]{revtex4}

\usepackage[T1]{fontenc}
\usepackage{amsmath,amssymb}
\usepackage{epsfig}
\usepackage{url}
\usepackage{subfigure}
\usepackage{graphicx}
\usepackage{grffile}
\usepackage[usenames,dvipsnames]{color}
\usepackage{slashed}
\usepackage{array}
\usepackage{multirow}
\usepackage[colorlinks,citecolor=blue]{hyperref}
\usepackage{color}
\usepackage{cancel}
\usepackage{gensymb}

\def\l {\lambda}

\def\bar {\overline}

\def\be {\begin{equation}}
\def\ee {\end{equation}}
\def\beq {\begin{equation}}
\def\eeq {\end{equation}}
\def\bea {\begin{eqnarray}}
\def\eea {\end{eqnarray}}

\def\bra {\langle}
\def\ket {\rangle}

\newcommand{\besub}{\begin{subequations}}
	\newcommand{\eesub}{\end{subequations}}

\def\beq{\begin{equation}}
\def\eeq{\end{equation}}
\def\barr{\begin{array}}
	\def\earr{\end{array}}

\begin{document}
	\title{Thermally corrected masses and freeze-in dark matter: a case study}

	\author{Nabarun Chakrabarty}
	\email{nabarunc@iitk.ac.in}
	\affiliation{Department of Physics, Indian Institute of Technology Kanpur, Kanpur, Uttar Pradesh-208016, India} 
	
	\author{Partha Konar}
	\email{konar@prl.res.in}
	\affiliation{Physical
		Research Laboratory, Ahmedabad - 380009, Gujarat, India}
	
	\author{Rishav Roshan}
	\email{rishav.roshan@gmail.com}
	\affiliation{Department of Physics, Kyungpook National University, Daegu 41566, Korea}
	
	\author{Sudipta Show}
	\email{sudipta@prl.res.in}
	\affiliation{Physical
		Research Laboratory, Ahmedabad - 380009, Gujarat, India\\
		Indian Institute of Technology, Gandhinagar - 382424, Gujarat, India}
	
	\begin{abstract} 
		If coupled \emph{feebly} to the Standard Model bath, a dark matter can evade the severe constraints from the direct search experiments. At the same time, such interactions help produce dark matter via the freeze-in mechanism. The freeze-in scenario becomes more interesting if one also includes the thermal masses of the different particles involved in the dark matter phenomenology. Incorporating such thermal corrections opens up the possibility of dark matter production via channels that remain kinematically disallowed in the standard freeze-in setup.  
		Motivated by this, we investigate such freeze-in production of the dark matter in a minimally extended $U(1)_{L_\mu-L_\tau}$ framework, which is also known to resolve the muon $g-2$ anomaly. Here, the role of the dark matter is played by a scalar with a non-trivial charge under the additional symmetry $U(1)_{L_\mu-L_\tau}$. The importance of incorporating the appropriate finite temperature corrections to freeze-in dynamics is aptly demonstrated in this study using the $U(1)_{L_\mu-L_\tau}$ scenario as a prototype.
		
	\end{abstract} 
	\maketitle


	\section{Introduction}
	\label{introduction}
	The existence of a non-luminous and a non-baryonic form of matter in the universe popularly known as \emph{dark matter} (DM)~\cite{Sofue:2000jx, Clowe:2006eq} has attracted the attention of the scientific community in the past several decades. Several vital pieces of evidence, like galactic rotation curves, gravitational lensing, and the anisotropies in the cosmic microwave background, have confirmed the presence of this form of matter in the universe. The vital information known about the DM is its relic density which is very accurately measured by the experiments studying the anisotropies in the cosmic microwave background  radiation~\cite{Planck:2018vyg, Hinshaw_2013}. Despite this vital information, the particle nature of the DM is still unknown. On the other hand, the Standard Model (SM) also fails to provide a particle that can be identified as a DM candidate. This suggests that physics beyond the SM is inevitable. 
	
	Depending on the interactions of the DM with the SM bath, several theories have been proposed in the literature that tries to explain the particle nature of the DM. 
	Among these theories, the most popular DM candidate is the \emph{weakly interacting massive particle } (WIMP)~\cite{Konar:2020wvl, Konar:2020vuu, Barman:2021qds, Chakrabarty:2021kmr, DuttaBanik:2020jrj, Borah:2020nsz, Bhattacharya:2019tqq, Borah:2019aeq,Baek:2008nz,Patra:2016shz,Biswas:2016yan}. Such a DM candidate thermalizes in the early Universe if the temperature of the thermal bath is above its mass. As the Universe cools down and the temperature of the plasma reaches below the mass of the DM mass, its abundance freezes out. Due to its not-so-small interactions with the SM particles, the WIMP type of dark matter is subjected to various experimental constraints. The experiments like LUX~\cite{LUX:2015abn, Akerib:2016vxi}, PANDA~\cite{Zhang:2018xdp}, XENON1T~\cite{Aprile:2018dbl}, {\color{blue}LUX-ZEPLIN (LZ)~\cite{LZ:2022ufs} }provide a stringent bound on the DM-nucleon scattering cross-sections. Such severe constraints can be easily evaded if the DM candidate is a \emph{feebly interacting massive particle} or FIMP~\cite{Hall:2009bx, Elahi:2014fsa, Bernal:2017kxu, Darme:2019wpd, Barman:2020plp, Barman:2021tgt, Datta:2021elq, Bhattacharya:2021jli, Konar:2021oye, Ghosh:2021wrk}. As the name itself suggest, this kind of DM interacts very feebly with the SM particles and hence cannot be tested in the experiments mentioned above. Due to the feeble interaction, a FIMP-type DM is slowly produced from the thermal bath, and once the number density of the bath particle that is responsible for its production becomes Boltzmann suppressed, the DM abundance freezes in.
	
	Recently, studies like~\cite{Darme:2019wpd, Konar:2021oye} have shown incorporating the thermal masses~\cite{PhysRevD.45.2933, Giudice:2003jh,Laine:2016hma, Biondini:2020ric} of the species involved in the DM production can open up the kinematically forbidden channels that are not allowed in the \emph{standard freeze-in }(SFI) scenarios. Here, it is assumed that the particle responsible for the production of the DM is not only a part of thermal plasma but also acquires a significant thermal mass at high temperatures due to its interactions with the thermal bath. As a result of the thermal corrections, the thermal mass of the species present in the plasma can be significantly different from their bare masses. 
	Consequently, even if a DM production channel is kinematically disallowed (forbidden) at $T = 0$ due to the hierarchy of masses between the bath particle and the dark matter state, that can change at a higher bath temperature. A $T \neq 0$ correction to the mass of the bath particle can flip the hierarchy, thereby kinematically opening up that particular production mode. In the specific context of DM production through the decay of a bath particle that picks up a thermal mass $M(T)$, the hierarchy $M(T)>2 M_{\text{DM}}$ is obtainable through non-zero temperature effects. 
	Such freeze-in production of the DM via forbidden channels is also known as \emph{forbidden freeze-in }(FFI). 
	An important study is \cite{Baker:2017zwx} that correlates cosmological phase transition, a phenomenon also driven by $T \neq 0$, to freeze-in in the presence of thermal corrections. Another one discussing thermal effects in DM and gravitational wave signatures is \cite{Ramazanov:2021eya}.
	
	Motivated by this, in the present article, we elucidate the idea of a
	spin-1 gauge boson receiving a thermally corrected mass in the early Universe and subsequently decaying to a DM particle. More precisely, we take the $U(1)_{L_\mu-L_\tau}$~\cite{He:1990pn,He:1991qd,Baek:2008nz,Biswas:2016yan,Biswas:2016yjr,Banerjee:2018eaf,Chun:2018ibr,Costa:2022oaa} framework for illustration. Unlike the standard $U(1)_{B-L}$ model~\cite{Bhattacharya:2019tqq,Konar:2021oye,Okada:2021nwo}, which offers a stable DM in the form of right-handed neutrino (RHN), the $U(1)_{L_\mu-L_\tau}$ scenario requires an additional scalar (singlet under the SM gauge symmetry) with a non-trivial $U(1)_{L_\mu-L_\tau}$ charge to explain the presence of the DM in the Universe. Assuming the DM interacts feebly with the bath particles, it can be produced from the decay of (i) the scalar responsible for the breaking of $U(1)_{L_\mu-L_\tau}$ symmetry, (ii) SM Higgs and, (iii) the massive gauge boson of $U(1)_{L_\mu-L_\tau}$ symmetry in the SFI scenario~\footnote{The production of DM from the scatterings can safely be ignored, as it remains suppressed in comparison to the decay.} if kinematically allowed, as was also discussed in \cite{ Biswas:2016yjr}. In this work, we aim to explore the deviation observed from the SFI scenario once the thermally corrected masses for the bath particles are taken into account. Besides explaining the dark matter, an $U(1)_{L_\mu-L_\tau}$ framework can simultaneously explain the discrepancy in the anomalous magnetic moment of muon $(g-2)$ from its SM prediction~\cite{Banerjee:2018eaf} and non-zero neutrino masses~\cite{Banerjee:2018eaf}. Keeping this in mind, we show that the present setup can accommodate a DM that can be produced via both SFI and FFI channels while also providing the solution for the discrepancy in the muon anomalous magnetic moment.
	
	The paper is organized as follows. The model is introduced in section~\ref{model}, and the various constraints deemed relevant are detailed in section~\ref{constraints}. 
	We compute the relevant thermal masses in section~\ref{DMPH}. And the same section also elaborates on the ensuing freeze-in phenomenology. Finally, the study is concluded in section~\ref{conclusion}.
	
	
	\section{The model}
	\label{model}
	
	We extend the SM gauge symmetry by an $U(1)_{L_\mu-L_\tau}$ symmetry where $L_\mu$ and $L_{\tau}$ represent the muon and tau lepton numbers, respectively. The fermionic content of the model includes the SM leptons and quarks together with three additional right-handed neutrinos ($N_e, N_\mu, N_{\tau}$). As suggested by the symmetry of the present scenario, the muon and tau carry a non-trivial charge under the  $U(1)_{L_\mu-L_\tau}$. The newly introduced RHNs are singlets under the SM gauge symmetry, while two of them carry 1 and $-1$ unit of $U(1)_{L_\mu-L_\tau}$, the third remains uncharged. We remind the readers that the addition of RHNs carrying non-zero $U(1)_{L_\mu-L_\tau}$ charges is necessary to cancel anomalies. The scalar sector of the setup is enhanced with a complex scalar ($S$) which is a singlet under the SM gauge symmetry but carries $1$ unit of $U(1)_{L_\mu-L_\tau}$ charge. We also introduce an additional scalar ($\phi$), a SM gauge singlet that plays the role of the DM. The stability of the DM is guaranteed by its non-trivial charge assignment under $U(1)_{L_\mu-L_\tau}$ symmetry. The gauge quantum numbers of the relevant fields are
	shown in Table~\ref{tab1} and Table~\ref{tab2}.
	\begin{center}
		\begin{table}[h!]
			\begin{tabular}{||c|c|c|c||}
				\hline
				\hline
				\begin{tabular}{c}
					Gauge\\
					Group\\ 
					\hline
					
					${\rm SU(2)}_{\rm L}$\\ 
					\hline
					${\rm U(1)}_{\rm Y}$\\ 
				\end{tabular}
				&
				
				\begin{tabular}{c|c|c}
					\multicolumn{3}{c}{Baryon Fields}\\ 
					\hline
					$Q_{L}^{i}=(u_{L}^{i},d_{L}^{i})^{T}$&$u_{R}^{i}$&$d_{R}^{i}$\\ 
					\hline
					$\textbf{2}$&$\textbf{1}$&$\textbf{1}$\\ 
					\hline
					$1/6$&$2/3$&$-1/3$\\ 
				\end{tabular}
				&
				\begin{tabular}{c|c|c}
					\multicolumn{3}{c}{Lepton Fields}\\
					\hline
					$L_{L}^{i}=(\nu_{L}^{i},e_{L}^{i})^{T}$ & $e_{R}^{i}$ & $N_{iR}$\\
					\hline
					$\textbf{2}$&$\textbf{1}$&$\textbf{1}$\\
					\hline
					$-1/2$&$-1$&$0$\\
				\end{tabular}
				&
				\begin{tabular}{c|c|c}
					\multicolumn{3}{c}{Scalar Fields}\\
					\hline
					$H$&$S$&$\phi$\\
					\hline
					$\textbf{2}$&$\textbf{1}$&$\textbf{1}$\\
					\hline
					$1/2$&$0$&$0$\\
				\end{tabular}\\
				\hline
				\hline
			\end{tabular}
			\caption{Particle contents and their
				charge assignments under the SM gauge group.}
			\label{tab1}
		\end{table}
	\end{center}
	\begin{center}
		\begin{table}[h!]
			\begin{tabular}{||c|c|c|c||}
				\hline
				\hline
				\begin{tabular}{c}
					Gauge\\
					Group\\ 
					\hline
					$U(1)_{L_\mu-L_\tau}$\\ 
					
				\end{tabular}
				&
				\begin{tabular}{c}
					\multicolumn{1}{c}{Baryon Fields}\\ 
					\hline
					$(Q^{i}_{L}, u^{i}_{R}, d^{i}_{R})$\\ 
					\hline
					$0$ \\ 
					
				\end{tabular}
				&
				\begin{tabular}{c|c|c}
					\multicolumn{3}{c}{Lepton Fields}\\ 
					\hline
					$(L_{L}^{e}, e_{R}, N_{e})$ & $(L_{L}^{\mu}, \mu_{R},
					N_{\mu})$ & $(L_{L}^{\tau}, \tau_{R}, N_{\tau})$\\ 
					\hline
					$0$ & $1$ & $-1$\\ 
					
				\end{tabular}
				&
				\begin{tabular}{c|c|c}
					\multicolumn{3}{c}{Scalar Fields}\\
					\hline
					$H$ & $S$ & $\phi$ \\
					\hline
					$0$ & $1$ & $n_{\mu \tau}$\\
				\end{tabular}\\
				\hline
				\hline
			\end{tabular}
			\caption{Particle contents and their
				charge assignments under $U(1)_{L_\mu-L_\tau}$.}
			\label{tab2}
		\end{table}
	\end{center}
	
	With an idea of particle content and their charges under the different symmetry groups, we now proceed to write their interactions. To begin with, we first write the kinetic terms for the additional fields,
	\begin{align}
	\mathcal{L}^{\text{KE}} &=\frac{i}{2}\sum_{i=e,\mu,\tau}\bar{N}_i\gamma^\delta D_\delta N_i+(D^\delta S)^\dagger(D_\delta S)+(D^\delta \phi)^\dagger(D_\delta \phi)-\frac{1}{4}F_{\mu\tau}^{\alpha\beta}F_{\mu\tau_{\alpha\beta}}&
	\label{kinetic_term}
	\end{align} 
	
	where $D_{\delta}=\partial_\delta+ig_{\mu\tau}Q_{\mu\tau}(Z_{\mu\tau})_\delta$ with $Q_{\mu\tau}$ representing  the charge and $Z_{\mu\tau}$ being the gauge boson of $U(1)_{L_\mu-L_\tau}$ symmetry. Finally $F_{\mu\tau}^{\alpha\beta}=\partial^\alpha Z_{\mu\tau}^\beta-\partial^\beta Z_{\mu\tau}^\alpha$. We also consider the pure $U(1)_{L_\mu-L_\tau}$ model where tree-level mixing of $Z_{\mu\tau}$ with the SM gauge bosons is absent. However, there is non-zero kinetic mixing at the one-loop level \cite{Hapitas:2021ilr} mediated by the leptons from the $\mu$- and $\tau$-families. Next, we write the Lagrangian involving the Yukawa interactions and masses of the additional fermions involved,
	\begin{align}
	\mathcal{L}=&-\frac{1}{2}h_{e\mu}(\bar{N}^c_e N_\mu+\bar{N}^c_\mu N_e)S^\dagger-\frac{1}{2}h_{e\tau}(\bar{N}^c_e N_\tau+\bar{N}^c_\tau N_e)S-\sum_{i=e,\mu,\tau}y_i\bar{L}_i \tilde{H}N_i \nonumber\\
	&-\frac{1}{2}M_{ee}\bar{N}^c_e N_e-\frac{1}{2}M_{\mu\tau}(\bar{N}^c_\mu N_\tau+\bar{N}^c_\tau N_\mu)S+h.c.
	\label{yukawa_term}
	\end{align}
	
	Finally, we write the most general scalar potential involving all the scalars in the present setup,
	\begin{align}
	V(H,S,\phi) =& -\mu^2_H H^\dagger H -\mu^2_S S^\dagger S
	+ \mu^2_\phi \phi^\dagger \phi
	+ \lambda_H (H^\dagger H)^2 + \lambda_S (S^\dagger S)^2
	+ \lambda_\phi (\phi^\dagger \phi)^2 \nonumber \\
	&
	+ \lambda_{HS} (H^\dagger H)(S^\dagger S)
	+ \lambda_{H\phi} (H^\dagger H)(\phi^\dagger \phi)  
	+ \lambda_{S\phi} (S^\dagger S) (\phi^\dagger \phi).
	\label{potential}
	\end{align}
	
	The scalar $S$ breaks the $U(1)_{L_\mu-L_\tau}$ symmetry once its CP even component develops a non-zero vacuum expectation value (vev) $v_{\mu\tau}$. The $U(1)_{L_\mu-L_\tau}$ gauge boson consequently obtains a non-zero mass, $m_{Z_{\mu\tau}}=g_{\mu\tau}v_{\mu\tau}$. The same breaking also results in an additional non-zero mixing that develops between the RHNs, as seen from Eq.~\ref{yukawa_term}. After the Electroweak Symmetry Breaking (EWSB), the Higgs doublet ($H$) also develops a non-zero vev $v = 246$ GeV. The scalars after the breaking of the gauge symmetry can be parameterized as,
	\begin{align}
	H =
	\begin{pmatrix}
	0\\
	\frac{1}{\sqrt 2}(v + h)
	\end{pmatrix},~~ S= \frac{1}{\sqrt 2}(v_{\mu\tau} + s ).
	\end{align}
	
	Subsequent to the EWSB, a non-zero $h-s$ mixing leads to the following mass terms,
	\begin{align}
	V \supset \frac{1}{2} \begin{pmatrix}
	h & s
	\end{pmatrix} \begin{pmatrix}
	\lambda_{H} \, v^2 & \lambda_{H S} \, v \, v_{\mu\tau} \\
	\lambda_{H S} \, v \, v_{\mu\tau} & 2 \lambda_S \, v_{\mu\tau}^2 
	\end{pmatrix} 
	\begin{pmatrix}
	h \\
	s 
	\end{pmatrix}.
	\label{scalars}  
	\end{align}
	The mass matrix is diagonalized using
	\begin{align}
	\begin{pmatrix}
	h \\
	s
	\end{pmatrix} = \begin{pmatrix}
	c_\theta & s_\theta \\
	-s_\theta & c_\theta 
	\end{pmatrix}
	\begin{pmatrix}
	h_1 \\
	h_2 
	\end{pmatrix}
	\label{scalar_mix}
	\end{align}
	with 
	\bea
	\tan{2\theta} &=& \frac{-2 \, \l_{H S} \, v \, v_{\mu\tau}}{\l_{H} \,  v^2 - 2 \, \l_S \, v_{\mu\tau}^2 }.
	\eea
	The mass eigenstates ($h_1,h_2$) then have masses
	\besub
	\bea
	m^2_{h_1,h_2} &=& \frac{1}{2} \Big[\big(\l_{H} v^2 + 2 \l_S v_{\mu\tau}^2\big) \pm \sqrt{(\l_H v^2 - 2 \l_S v_{\mu\tau}^2\big)^2 + 4 \l_{\phi S}^2 v^2 v_{\mu\tau}^2}\Big]. 
	\eea
	\eesub
	The various model parameters are expressible in terms of the physical quantities as follows: 
	\besub
	\bea
	\l_H &=& \frac{m^2_{h_1}c^2_\theta + m^2_{h_2}s^2_\theta}{v^2}, \label{lH}\\
	\l_S &=& \frac{m^2_{h_1}s^2_\theta + m^2_{h_2}c^2_\theta}{v_{\mu\tau}^2}, \label{lS}\\
	\l_{HS} &=& \frac{2(m^2_{h_1} - m^2_{h_2})s_\theta c_\theta}{v v_{\mu\tau}}\label{lHS}.
	\eea
	\eesub
	
	Finally, after both the symmetries are broken, the dark matter mass can be expressed as,
	\bea
	m_\phi^2 &=& \mu^2_\phi + \frac{1}{2}\l_{H\phi} v^2 + \frac{1}{2}
	\l_{S\phi} v^2_{\mu\tau}.
	\eea 
	It is convenient to describe a framework in terms of vevs, physical masses, and mixing angles. We demand $m_{h_1}$ = 125 GeV and tag $\{g_{\mu\tau},v_{\mu\tau},m_{h_2},m_\phi,s_\theta,\l_{H\phi},\l_{S\phi} \}$ as the free parameters of the 
	scalar sector.

	\section{Constraints and additional issues}
	\label{constraints}
	
	We discuss in this section the various constraints on this model, as well as the predictions of neutrino mass and the muon anomalous magnetic moment in this model.
	
	\subsection{Neutrino scattering experiments}
	
	Gauging the $U(1)_{L_\mu-L_\tau}$ symmetry leads to severe constraints
	from neutrino trident production, that is, $\nu_\mu (\bar{\nu}_\mu) + N \to \nu_\mu (\bar{\nu}_\mu) + \mu^+ \mu^- + N$. Here $N$ denotes a heavy nucleus. Given the good agreement of the observed results with the SM for this process reported by CHARM-II~\cite{CHARM-II:1990dvf} and CCFR~\cite{CCFR:1991lpl, Altmannshofer:2014pba}, the parameter space in the presence of a new neutral gauge boson gets seriously restricted. A relatively new probe is the coherent elastic neutrino-nucleus scattering (CE$\nu$NS) process that, at the amplitude level, looks like $\nu N \to \nu N$. CE$\nu$NS has recently been measured by the COHERENT collaboration~\cite{doi:10.1126/science.aao0990,PhysRevD.95.115028,LIAO201754}. The $Z_{\mu\tau}$ in the gauged $U(1)_{L_\mu-L_\tau}$ model enters
	CE$\nu$NS through kinetic mixing with the SM gauge bosons and thereby gets constrained. In fact, the BOREXINO~\cite{PhysRevD.95.055006,PhysRevD.98.015005} process studies the same scattering process as COHERENT, however, using solar neutrinos. The impact of all the constraints is conveniently depicted in the 
	$m_{Z_{\mu\tau}}-g_{\mu\tau}$ plane in Fig. \ref{f:const}.

	\subsection{LHC constraints}
	$Z \to 4\mu$ searches by ATLAS~\cite{PhysRevLett.112.231806} and CMS~\cite{CMS:2012bw} constrains the gauge sector of the model and rules out a portion of the $m_{Z_{\mu\tau}}-g_{\mu\tau}$ plane as shown in Fig. \ref{f:const}. On another side, an $h-s$ mixing as defined by Eq.(\ref{scalar_mix}) implies that the tree-level couplings of $h_1$ with the SM fermions and gauge bosons scale by a factor of $c_\theta$ \emph{w.r.t.} the corresponding SM ones. This subjects the mixing angle $\theta$ to Higgs signal strength constraints and other exclusion limits from the LHC~\cite{ATLAS:2016neq, ATLAS:2019cid}. We adopt $|s_\theta| < 0.1$ in this work to comply with all such constraints. Finally, the reported upper limit on the invisible branching ratio of the 125 GeV Higgs puts a limit on BR$(h \to \phi \phi^\dagger)$ whenever kinematically allowed. However, this constraint is almost trivially satisfied in this setup given a \emph{feeble} $h-\phi-\phi^\dagger$ interaction strength dictated by the freeze-in dynamics.
	
	\subsection{Dark matter constraints}
	
	We demand that the freeze-in relic density predicted by this model
	must entirely account for the observed relic of the universe. The Planck experiment~\cite{Planck:2018vyg} has reported
	\begin{align}
	\Omega_{\text{DM}} h^2 = 0.120 \pm 0.001.
	\end{align}
In addition, any dark matter model must abide by the limits imposed by the various direct detection experiments mentioned before and the most stringent constraint currently comes from LZ. Direct detection scatterings in this model are mediated by gauge bosons through kinetic mixing as well as through the scalars $h_1,h_2$. It is, however, much easier to evade such constraints in a freeze-in framework such as the present scenario. This is because demanding the observed relic density through the freeze-in dynamics would necssitate that couplings of the DM to $h_1,h_2$ are tiny and the charge $n_{\mu\tau}$ is accordingly small. The direct detection cross sections are thus expected to be below the stipulated bound.

\subsection{Muon $g-2$} 
	
\begin{figure}[htb!]
		\includegraphics[height=8cm,width=10cm]{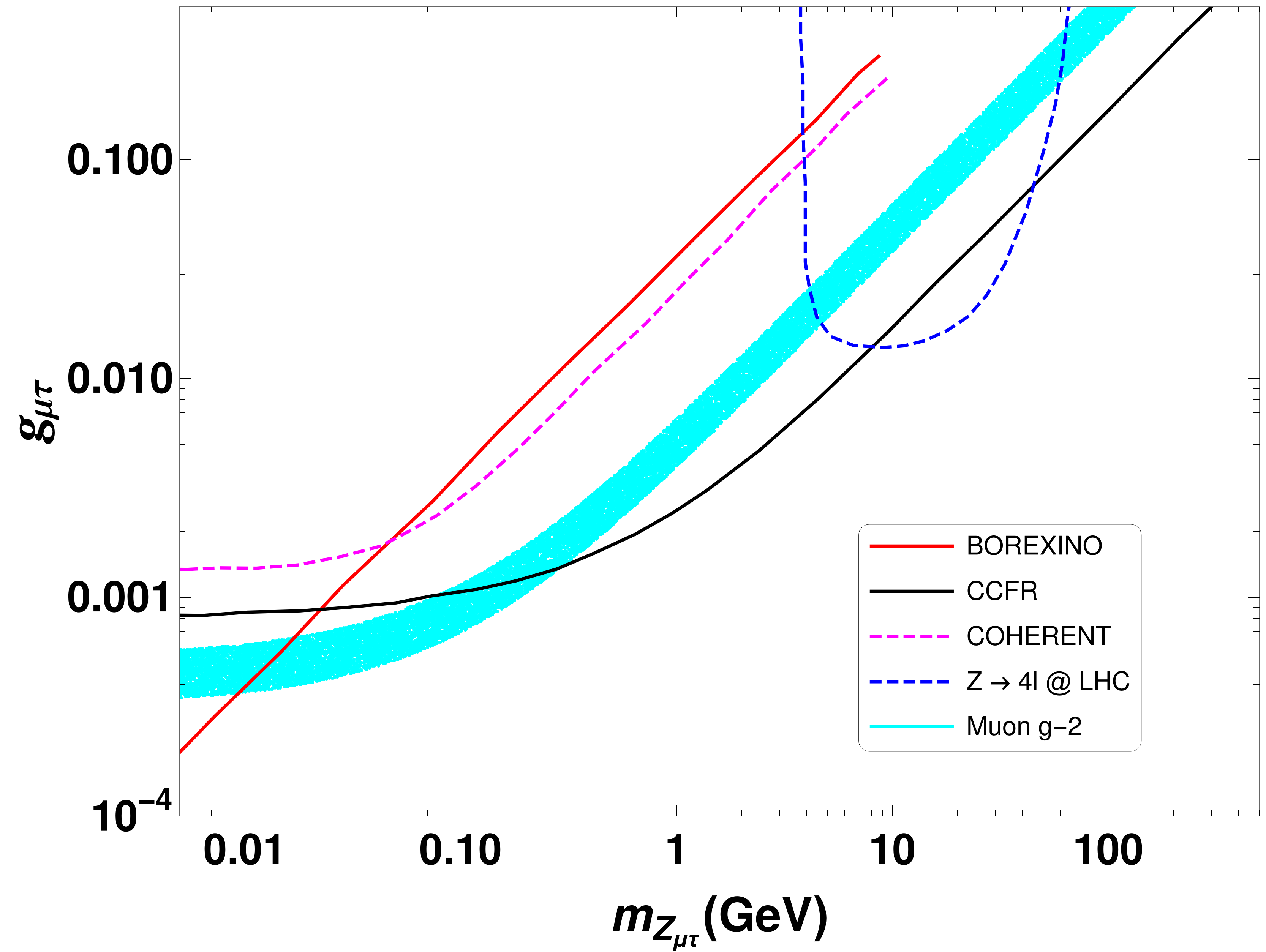}
		\caption{The impact of the various experimental constraints on the present model. The regions to the left of the red, magenta, and black curves are ruled out by BOREXINO, COHERENT, and CCFR. And the region bound by the blue curve is ruled out by the $Z \to 4 l$ searches at the LHC. The cyan band is the region compatible with the 2$\sigma$ limit of muon $g-2$ as quoted in Eq.\ref{gmt_limit}.}
		\label{f:const}
	\end{figure}

	The dominant contribution to $\Delta a_\mu$ in this model comes from the 1-loop amplitude mediated by the $Z_{\mu\tau}$ gauge-boson\footnote{The contribution coming from $N_i$ is suppressed and has been neglected.}. This contribution can be expressed as~\cite{Gninenko:2001hx, Baek:2001kca}
	\begin{align}
	\Delta a^{Z_{\mu\tau}}_\mu =\frac{g^2_{\mu\tau}}{4 \pi^2} \int_0^1 dx \frac{x (1-x)^2}{(1-x)^2 + r x},
	\end{align}
	where $r = (m_{Z_{\mu\tau}}/m_\mu)^2$. Following the announcement of the FNAL~\cite{Muong-2:2021ojo} results on muon $g-2$, a combined measurement of the discrepancy is
	\begin{align}
	a^{\text{exp}}_\mu - a^{\text{SM}}_\mu =
	(2.51 \pm 0.59) \times 10^{-9}\label{gmt_limit}.
	\end{align}
	An inspection of Fig. \ref{f:const} reveals that apart from the stretch around $m_{Z_{\mu\tau}} \in$ [10 MeV,300 MeV], the parameter space compatible with the observed $\Delta a_\mu$ is almost entirely ruled out by the neutrino scattering experiments.
	
	\subsection{Neutrino mass}
	
	Generation of neutrino mass in the gauged $U(1)_{L_\mu-L_\tau}$ model occurs through Type-I seesaw~\cite{Minkowski:1977sc,GellMann:1980vs,Mohapatra:1979ia,PhysRevD.22.2227,Schechter:1981cv} and has been discussed in detail in~\cite{Patra:2016shz, Biswas:2016yan, Banerjee:2018eaf}. The light neutrino mass matrix has the familiar Type-I form
	\begin{align}
	M_\nu = -M_D M^{-1}_R M^T_D\label{type1}.
	\end{align}
	Here, $M_D$ and $M_R$ refer to the Dirac and Majorana mass matrices. Following the spontaneous breaking of the gauge symmetry of the model, one derives
	\besub
	\bea
	M_R &=& \begin{pmatrix}
		M_{ee} & \frac{1}{\sqrt{2}} h_{e\mu} v_{\mu\tau}  & 
		\frac{1}{\sqrt{2}} h_{e\tau} v_{\mu\tau}
		\\
		\frac{1}{\sqrt{2}} h_{e\mu} v_{\mu\tau} &
		0 & M_{\mu\tau} \\
		\frac{1}{\sqrt{2}} h_{e\tau} v_{\mu\tau} &
		M_{\mu\tau} & 0
	\end{pmatrix}, \\
	M_D &=& \frac{1}{\sqrt{2}}v \times \text{diag}(y_e,y_\mu,y_\tau).
	\eea
	\eesub
	We refer the reader to~\cite{Patra:2016shz, Biswas:2016yan, Banerjee:2018eaf} for details of fitting the neutrino data. A similar approach is adopted for this study.

	\section{Freeze-in production and the impact of thermal corrections }
	\label{DMPH}

\begin{figure}[htb!]
		\includegraphics[height=4cm,width=15cm]{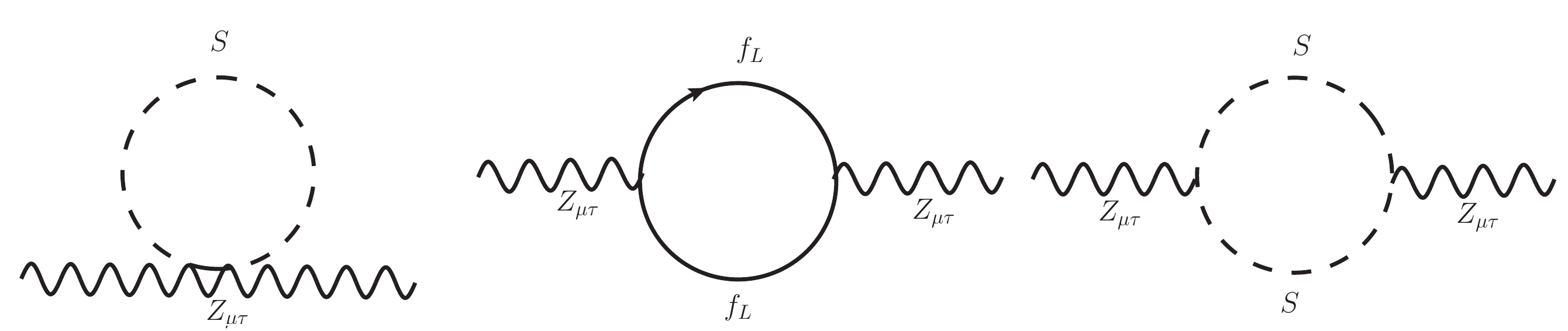}
		\caption{The one loop diagrams contributing to the thermal mass of gauge boson $Z_{\mu\tau}$.}
		\label{Ther_corr_gauge_boson}
	\end{figure}

	This section outlines the impact of $T\neq 0$ on the masses of particles in this model. 
	The formalism we follow is elaborately discussed in the review~\cite{Laine:2016hma}. Henceforth,
	the thermal correction to the mass of a particle $P$ will be denoted by $\delta m^2_P(T)$ and its thermally corrected mass by $M_P(T)$. One then notes $M_P(T) = \sqrt{m^2_P + \delta m^2_P(T)}$ where $m_P$ is the mass for $T=0$. We first discuss the correction to the $Z_{\mu\tau}$ mass.
	The contribution coming from a complex scalar carrying a charge $q_S$ (see left panel of Fig.~\ref{Ther_corr_gauge_boson}) is given by
	\begin{align}
	\delta M^2_{Z_{\mu\tau}}(T)\big|_S =\frac{1}{3}
	q^2_S g^2_{\mu\tau} T^2.
	\end{align}
	We add here that only the longitudinal component of a gauge boson receives thermal corrections. Similarly, the contribution coming from a chiral fermion (see right panel of Fig.~\ref{Ther_corr_gauge_boson}), say $f_L$ carrying 
	$q_f$ charge reads
	\begin{align}
	\delta M^2_{Z_{\mu\tau}}(T)\big|_{f_L} = \frac{1}{6}
	(q_f)^2 g^2_{\mu\tau} T^2.
	\end{align}
	
It can be checked that possible $\mathcal{O}(T^2)$ contributions to $B-Z_{\mu\tau}$ and $W^3-Z_{\mu\tau}$ mixings vanish on account of the opposite $U(1)_{L_\mu-L_\tau}$ charges of the $\mu$- and $\tau$-family.
	Thus, summing up the contributions coming from all relevant fields in the model, one obtains
	\begin{align}
	\delta M^2_{Z_{\mu\tau}}(T)= \Big(\frac{5}{3} + n^2_{\mu\tau}\Big)g^2_{\mu\tau} T^2.
	\end{align}

	As detailed in the previous sections, the DM $\phi$ has \emph{feeble} interactions with the scalars as well as with the gauge field $Z_{\mu\tau}$. The smallness of such interaction strengths implies that the interaction rate of DM remains smaller than the Hubble expansion rate throughout the thermal course of the universe. Consequently,
	the DM $\phi$ is never in equilibrium with the thermal bath and is injected into the thermal plasma via annihilations and decays of other particles. This is the \emph{freeze-in} mechanism in a nutshell. Of these, the dominant contribution comes from the decay since the annihilations typically undergo suppressions by propagators and additional couplings. The impact of the annihilations is hence neglected in this study hereafter.  Also, since the DM $\phi$ does not enter the thermal bath at any point in its cosmological history, it is \emph{cold}. Therefore, thermal corrections to the DM mass itself become negligible in this setup.
	
\begin{figure}[htb!]
		\includegraphics[height=4.5cm,width=10cm]{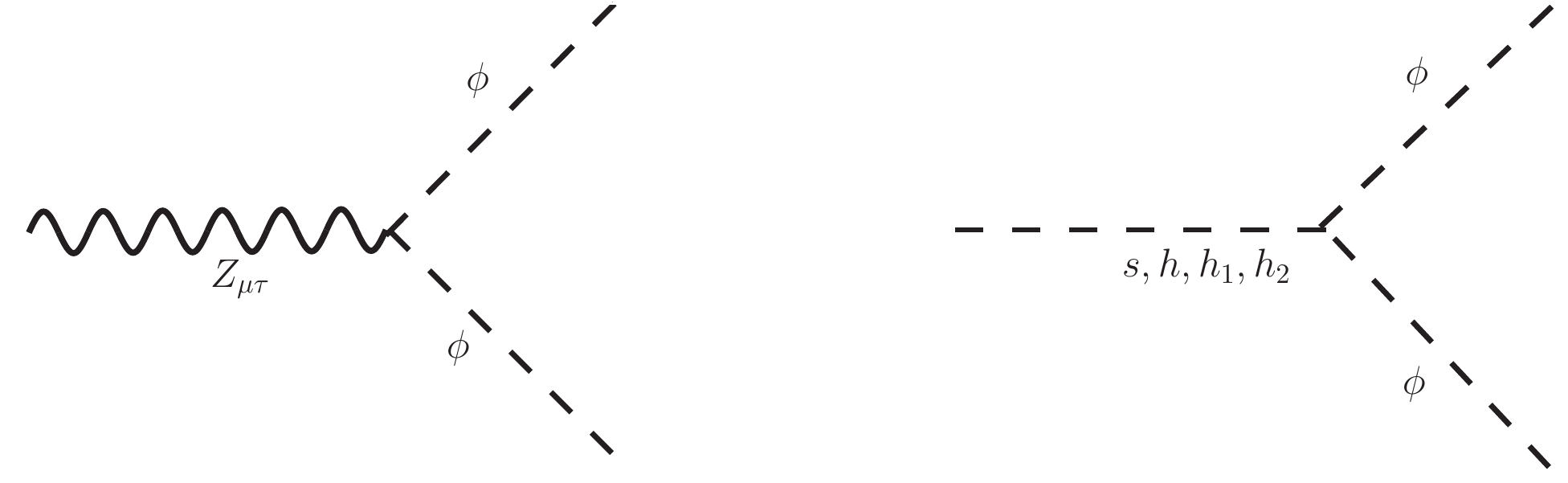}
		\caption{Decays responsible for the dark matter production.}
		\label{FI_decay}
	\end{figure}

	We remind here that some previous studies~\cite{Biswas:2016yjr} have looked at the freeze-in dynamics for the $U(1)_{L_\mu-L_\tau}$ model in detail. As mentioned earlier, we shall refer to the standard picture that has emerged from such studies as "standard freeze-in" or SFI. Our goal in this study is to demonstrate the deviation from SFI when thermal corrections to the masses of both the decaying particle and the DM are taken. We assume that all the decaying particles (see Fig.~\ref{FI_decay}), i.e., the two scalars and $Z_{\mu\tau}$, are throughout in thermal equilibrium. 
	
	The Boltzmann equation predicting the DM yield is then given by
	\begin{align}
	\frac{d Y_\phi}{d x} = \frac{1}{H_0 x} \bigg[ <\Gamma_{Z_{\mu\tau} \to \phi \phi^\dagger}>(x) Y^{\text{eq}}_{Z_{\mu\tau}}
	+ \sum_{i=1,2} <\Gamma_{S_i \to \phi \phi^\dagger}>(x) Y^{\text{eq}}_{S_i} \bigg].
	\end{align}
	Here $x = \frac{m_\phi}{T}$ with $T$ and $H_0$ = $1.67 \sqrt{g_*} \frac{T^2}{M_{\text{Pl}}}$ denoting the temperature and the expansion rate
	of the universe respectively. In addition, $Y_\phi = \frac{n_\phi}{s}$ refers to the comoving number density of the
	DM with s being the entropy density. $Y_i^\text{eq}$ signifies
	the equilibrium densities with $i=Z_{\mu\tau},S_1,S_2$.
	Here, in theory, $S_1, S_2$ generically denote the two neutral scalars. It is pointed out that $\{S_1,S_2\}$ respectively coincide with $\{h,S\}$ and 
	$\{h_1,h_2\}$ before and after the spontaneous breakdown of the gauge symmetry. The thermal decay width for $A \to B ~C$ reads 
	\begin{align}
	\bra\Gamma_{A \to B~C}\ket(x) = \frac{K_1(x)}{K_2(x)}~\Gamma_{A \to B~C},
	\end{align}
	where $K_n(x)$ is the $n$th order modified Bessel function.  The late-time DM yield $Y_\phi(x_\infty)$
	is calculated by solving the Boltzmann equation. The DM relic density is then obtained using 
	\begin{align}
	\Omega_{\phi} h^2 = 2.75 \times 10^{8}~m_\phi Y_\phi(x_\infty). 
	\end{align}
We mention here we have neglected the effect of kinetic mixing (a $T = 0$ phenomenon) on freeze-in production as it is radiatively suppressed~\cite{Hapitas:2021ilr} and therefore small. It is reminded that the parameters controlling the interaction strengths of $\phi$ and ultimately $Y_\phi(x_\infty)$ are $\l_{H\phi},\l_{S\phi},g_{\mu\tau}$ and $n_{\mu\tau}$. For a clearer understanding of the interplay of the different thermal masses involved, we divide the subsequent analysis into Scenario A: $\l_{H\phi},\l_{S\phi} < < n_{\mu\tau}g_{\mu\tau}$ and Scenario B: $\l_{H\phi},\l_{S\phi} \sim n_{\mu\tau}g_{\mu\tau}$. We further take $g_{\mu\tau} = 5 \times 10^{-4}$ and $v_{\mu\tau}$ = 80 GeV for this case which corresponds to the tree level mass 
	$M_{Z_{\mu\tau}}$ = 0.04 GeV following the spontaneous breaking of $U(1)_{L_\mu - L_\tau}$. Moreover, this choice predicts
	$\Delta a_\mu = 1.45 \times 10^{-9}$ and thus is consistent with the latest 2$\sigma$ experimental limit of Eq.(\ref{gmt_limit}).
	Explicit verifications establish that the lifetime corresponding to the $Z_{\mu\tau} \to \nu_\mu \bar{\nu}_\mu, \nu_\tau \bar{\nu}_\tau$ decays is smaller than the age of the Universe by several orders of magnitude for the said choice of $g_{\mu\tau}$. Thus $Z_{\mu\tau}$ is not cosmologically stable and thus does not contribute to the relic density. Finally, we would also like to emphasize that the interaction strength of $Z_{\mu\tau}$ with the SM fermions is large enough to keep it in equilibrium in the early universe.
	
	\subsection{$\l_{H\phi},\l_{S\phi} < < n_{\mu\tau}g_{\mu\tau}$} 
	This limit entails that $\Gamma_{S_i \to \phi \phi^\dagger} < < \Gamma_{Z_{\mu\tau} \to \phi \phi^\dagger}$ and hence DM is dominantly produced by the $Z_{\mu\tau}$ decay. In order to study the impact of thermal corrections on DM production, we propose $m_\phi$ = 1 GeV, 30 MeV, and 25 MeV as benchmark values. The rationale behind choosing such values will become clear in the  
		ensuing discussion.

\begin{figure}[htb!]
		\includegraphics[height=6.5cm,width=8cm]{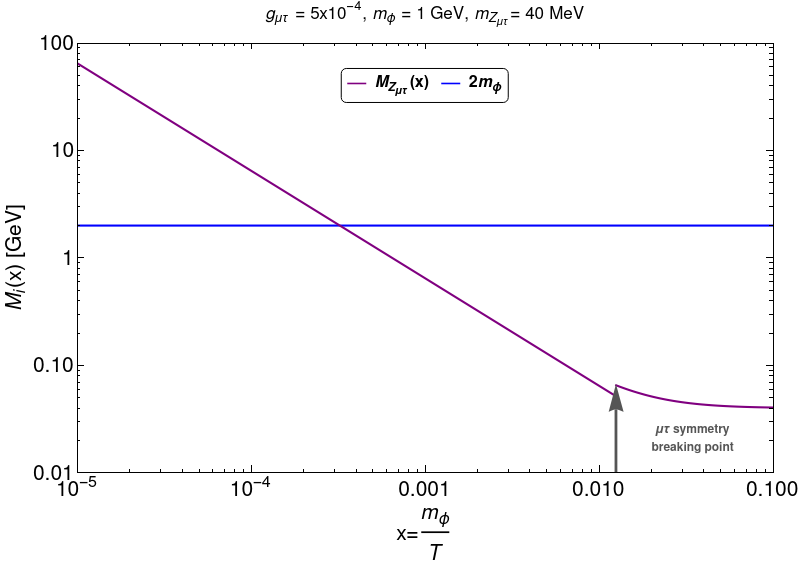}
		\includegraphics[height=6.5cm,width=8cm]{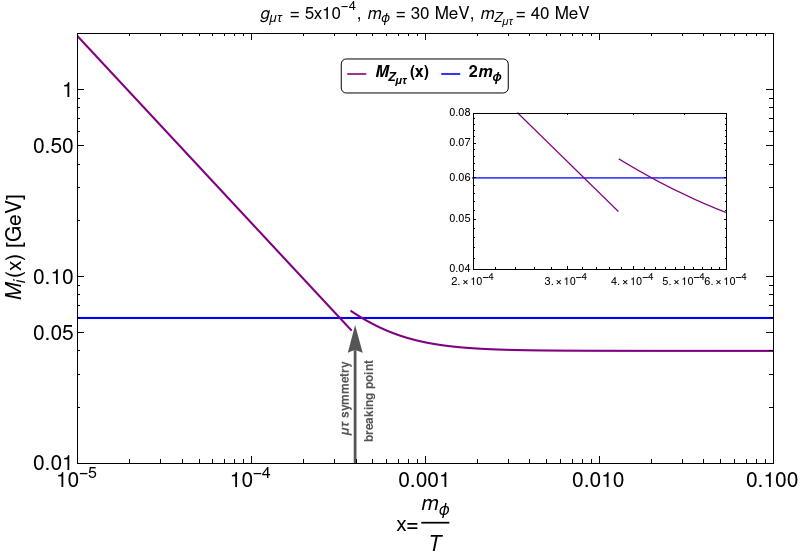} 
		\includegraphics[height=6.5cm,width=8cm]{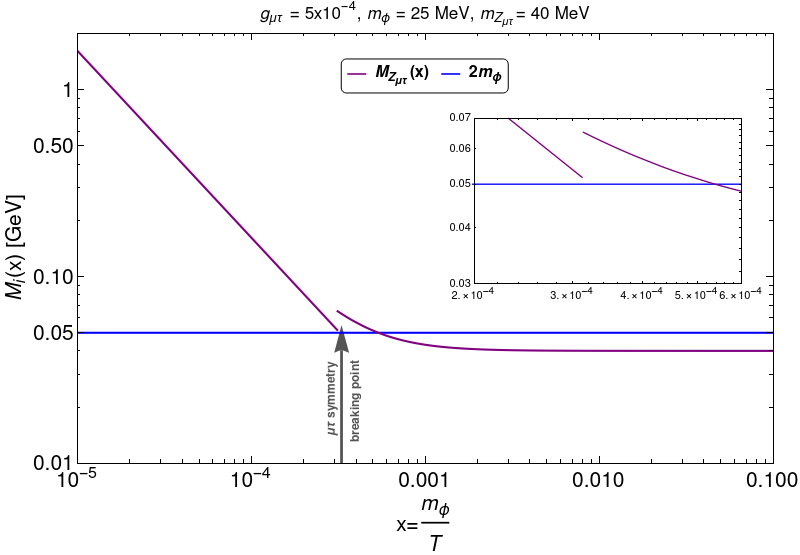} 
		\caption{Variation of thermal masses of the extra gauge boson$({Z_{\mu\tau}})$ (purple) with the dimensionless variable $x=\frac{m_\phi}{T}$ for three different values of DM masses. The horizontal blue line corresponds to  $2m_\phi$ in all the panels. }
		\label{f:mass_A}
	\end{figure}

	The thermally corrected mass of $Z_{\mu\tau}$ for $T > v_{\mu\tau}$ reads $M_{Z_{\mu\tau}}(T) \simeq \sqrt{\frac{5}{3}} g_{\mu\tau}T$ since $n_{\mu\tau} \ll 1$ for freeze-in.  
		It is once again reminded that while the mass of $Z_{\mu\tau}$ entirely comes from thermal corrections for $T > v_{\mu\tau}$, the DM 
		$\phi$ does get a bare mass equalling $\mu_\phi$ from the scalar potential in the said temperature range. Fig.~\ref{f:mass_A} shows the variation of $M_{Z_{\mu\tau}}(x)$ for the chosen values of $m_\phi$. At a very high temperature, say $T_{\text{initial}} = 10^5$ GeV ($x_{\text{initial}} = 10^{-5}$ for $m_\phi$ = 1 GeV), one finds
		$M_{Z_{\mu\tau}}(T_{\text{initial}})$ = 64.55 GeV. However, this mass gap diminishes with decreasing $T$ (increasing $x$) and a crossover is observed at $T=T_1^{\text{cr}}$ ($x=x_1^{\text{cr}}$) obtainable through $M_{Z_{\mu\tau}}(T_1^{\text{cr}}) = 2m_{\phi}$,
		beyond which $M_{Z_{\mu\tau}}(T) < 2m_{\phi}$. We hereafter refer to this as the \emph{first crossover}. Using the expressions for $M_{Z_{\mu\tau}}(T)$ given above, one derives $T_1^{\text{cr}} = \sqrt{\frac{12}{5}} \frac{m_\phi}{g_{\mu\tau}}$. This crossover is seen to happen for $m_{\phi}$ = 1 GeV and 30 MeV. Table \ref{tab:crT_A} displays the corresponding $T_1^{\text{cr}}$ values.
		As an example, $m_{\phi}$ = 1 GeV predicts $T_1^{\text{cr}} \simeq 3.098 \times 10^{3}$ GeV. On the other hand, decreasing $m_\phi$  accordingly postpones the crossover. For instance, as $m_\phi$ is lowered from 1 GeV to 30 MeV, $T_1^{\text{cr}}$ proportionately decreases from $\simeq 3.098 \times 10^{3}$ GeV to 92.95 GeV.

	We next come to discuss the role of  $v_{\mu\tau}$ in this scenario. An inspection of Table \ref{tab:crT_A} reveals that $T_1^{\text{cr}} > v_{\mu\tau}$ = 80 GeV for $m_\phi$ = 1 GeV and 30 MeV. The spontaneous breaking of $U(1)_{L_\mu-L_\tau}$ takes place at $T = v_{\mu\tau}$ thereby generating a squared mass equalling $g^2_{\mu\tau} v^2_{\mu\tau}$ for $Z_{\mu\tau}$. The thermally corrected mass for the same therefore shows a kink at $T = v_{\mu\tau}$, as can be seen in Fig.~\ref{f:mass_A}. And this kink opens up the possibility of having $M_{Z_{\mu\tau}}(T = v_{\mu\tau}) > 2m_{\phi}$ for a second time during the thermal evolution of this scenario. We compute the $Z_{\mu\tau}$ thermal mass at this symmetry-breaking threshold in Table \ref{tab:crT_A} and discover that this indeed happens in case of $m_{\phi}$ = 30 MeV. Moreover, at some $T_2^{\text{cr}} < v_{\mu\tau}$, one again might encounter 
		$M_{Z_{\mu\tau}}(T_2^{\text{cr}}) = 2m_{\phi}$ for a second time. This is referred to here as the \emph{second crossover} with the corresponding temperature being 
		$T_2^{\text{cr}} = \sqrt{\frac{12 m^2_\phi - 3 g_{\mu\tau}^2 v_{\mu\tau}^2}{5 g_{\mu\tau}^2}}$. We mention here again that all crossover possibilities and the corresponding temperatures and $x$-values are summarised in Table \ref{tab:crT_A} for each $m_\phi$.
	
	We can take a stock at this point. Each $m_\phi$ value entails a qualitative behavior distinct from the others. For $m_\phi$ = 1 GeV, the first crossover indeed occurs following which $M_{Z_{\mu\tau}}(T)$ becomes smaller than $2m_{\phi}$ at all later times, or, at all lower temperatures. Even the kink at $T=v_{\mu\tau}$ does not flip the hierarchy between $M_{Z_{\mu\tau}}(T)$ and 
		$m_\phi$ immediately preceding $T = v_{\mu\tau}$. Hence, a second crossover is ruled out for this DM mass. On the other hand, following the first crossover in case of $m_\phi$ = 30 MeV, the kink at the symmetry breaking temperature again leads to $M_{Z_{\mu\tau}}(T)$ $ > 2m_{\phi}$. And the second crossover also takes place shortly after that. Lastly, for $m_\phi$ = 25 MeV, $M_{Z_{\mu\tau}}(T) > 2m_{\phi}$ is maintained all the way from $T > > v_{\mu\tau}$ to $T = T^{\text{cr}}_2$ following which the mass hierarchy flips.
	
	We would like to comment on the representativeness of the chosen benchmarks. Any horizontal line corresponding to a given $2m_\phi$ value that cuts the $M_{Z_{\mu\tau}}(x)$ curve just once for  $x < \frac{m_\phi}{v_{\mu\tau}}$ shares the same qualitative features as BP1. Similarly, the $2m_\phi$ line that cuts the $M_{Z_{\mu\tau}}(x)$ curve thrice, i.e., at $x < \frac{m_\phi}{v_{\mu\tau}}$, $x = \frac{m_\phi}{v_{\mu\tau}}$ and $x > \frac{m_\phi}{v_{\mu\tau}}$ would be qualitatively similar to BP2. Finally, the $2m_\phi$ line that cuts the $M_{Z_{\mu\tau}}(x)$ curve once for $x > \frac{m_\phi}{v_{\mu\tau}}$ is the same as BP3 qualitatively. 
	
\begin{table}[t]
	\centering
	\begin{tabular}{|l|c|c|c|c|c|c|}
		\hline
		Benchmark & $m_\phi$ 	& First  	&  $T_1^{\text{cr}}~(x_1^{\text{cr}})$ 	& $m_{X}(T = v_{\mu\tau})$ &  Second   & $T_2^{\text{cr}}~(x_2^{\text{cr}})$   \\     
		Points         &  	&  crossover		&   		&  	  &  	crossover  	 & 				\\     \hline  \hline
		BP1 & 1 GeV & Yes 		&  3098.39 GeV ($0.00032 $)  		& 0.065 GeV 	  &    No	 &  	-		\\     \hline
		BP2 & 30 MeV & Yes		& 92.95 GeV ($0.00032$) 		& 0.065 GeV	  &    Yes		 &  69.28 GeV ($0.00043$)	\\     \hline
		BP3 & 25 MeV & No		&  - 		& 0.065 GeV 	  &    	Yes	 &  	46.48 GeV ($0.00054$)		\\     \hline
	\end{tabular}
	\caption{Benchmark Points considered in the analysis. }
\label{tab:crT_A}
\end{table}

\begin{figure}[htb!]
		\includegraphics[height=6.5cm,width=8cm]{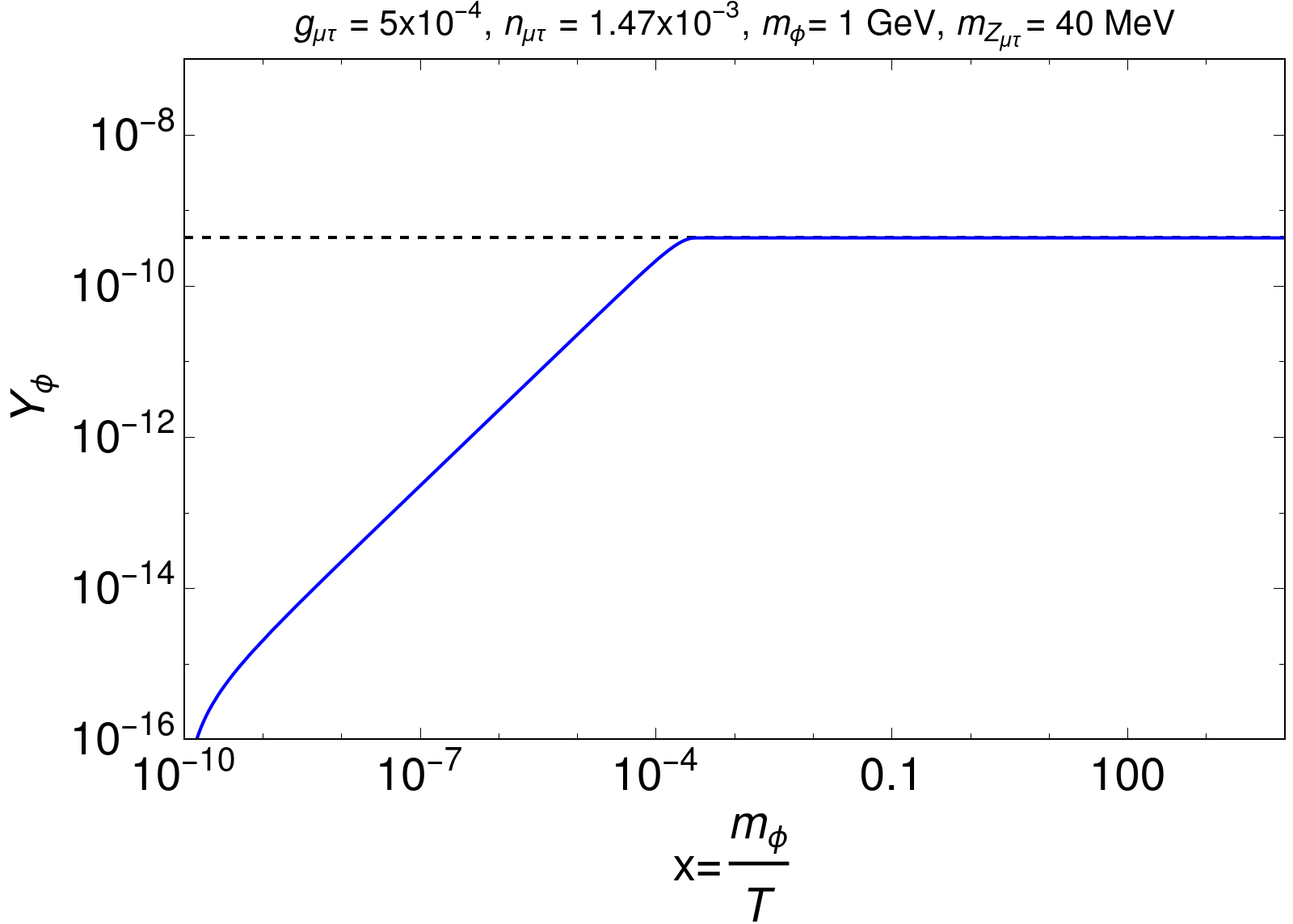}
		\caption{Evolution of the DM comoving number density {\color{blue}(in blue)} as a function of $x = \frac{m_\phi}{T}$ for $m_\phi$ = 1 GeV when DM production from scalar decay is negligible. The dashed horizontal line corresponds to the value of DM abundance ($Y_\phi$) for which the observed relic density is satisfied.}
		\label{yphi_wos_1}
	\end{figure}
\begin{figure}[htb!]
		\includegraphics[height=6.5cm,width=8cm]{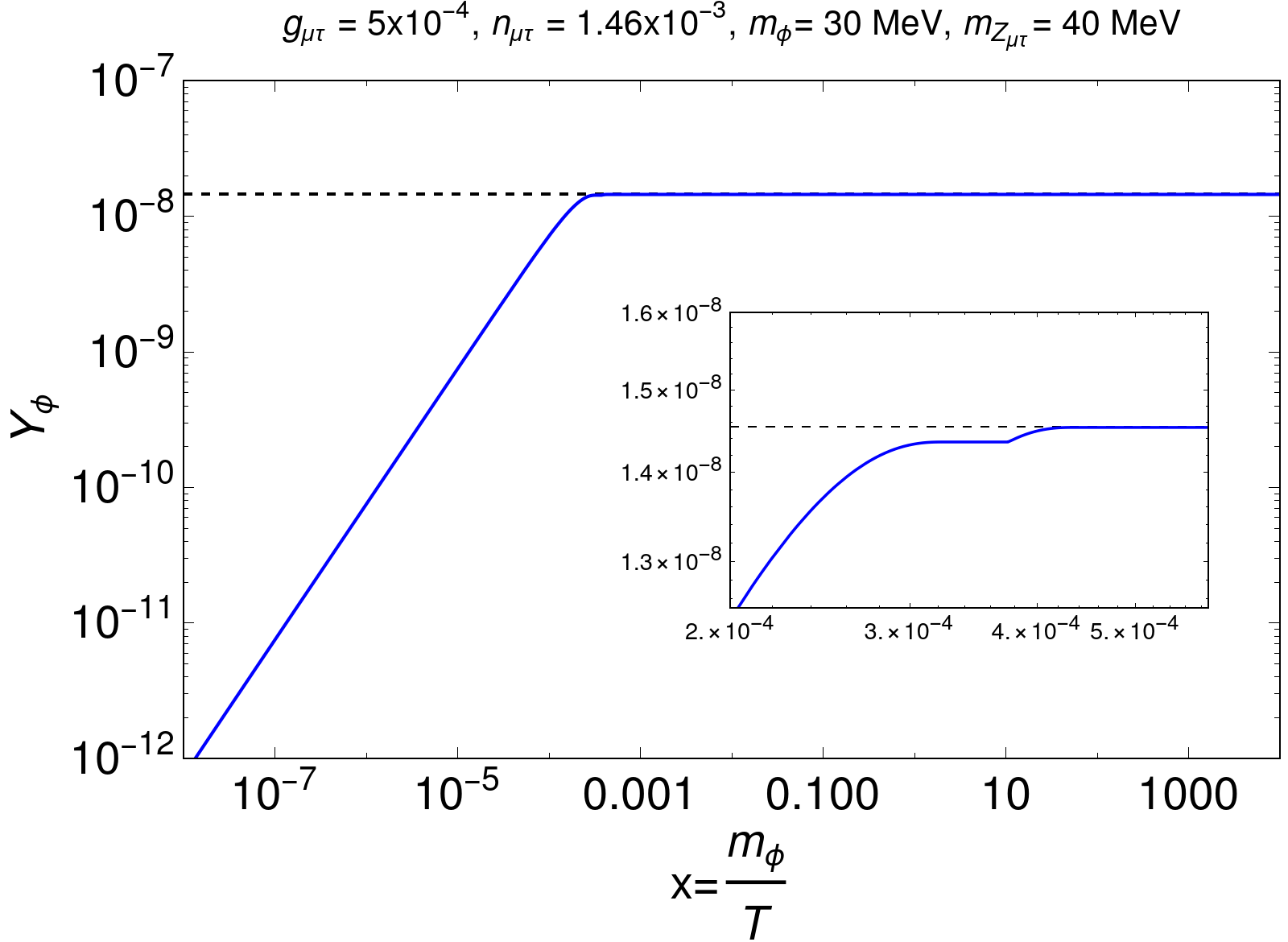}
		\caption{Evolution of the DM comoving number density {\color{blue}(in blue)} as a function of $x = \frac{m_\phi}{T}$ for $m_\phi$ = 30 MeV when DM production from scalar decay is negligible. The dashed horizontal line corresponds to the value of DM abundance ($Y_\phi$) for which the observed relic density is satisfied.}
		\label{yphi_wos_p030}
	\end{figure}
\begin{figure}[htb!]
		\includegraphics[height=6.5cm,width=8cm]{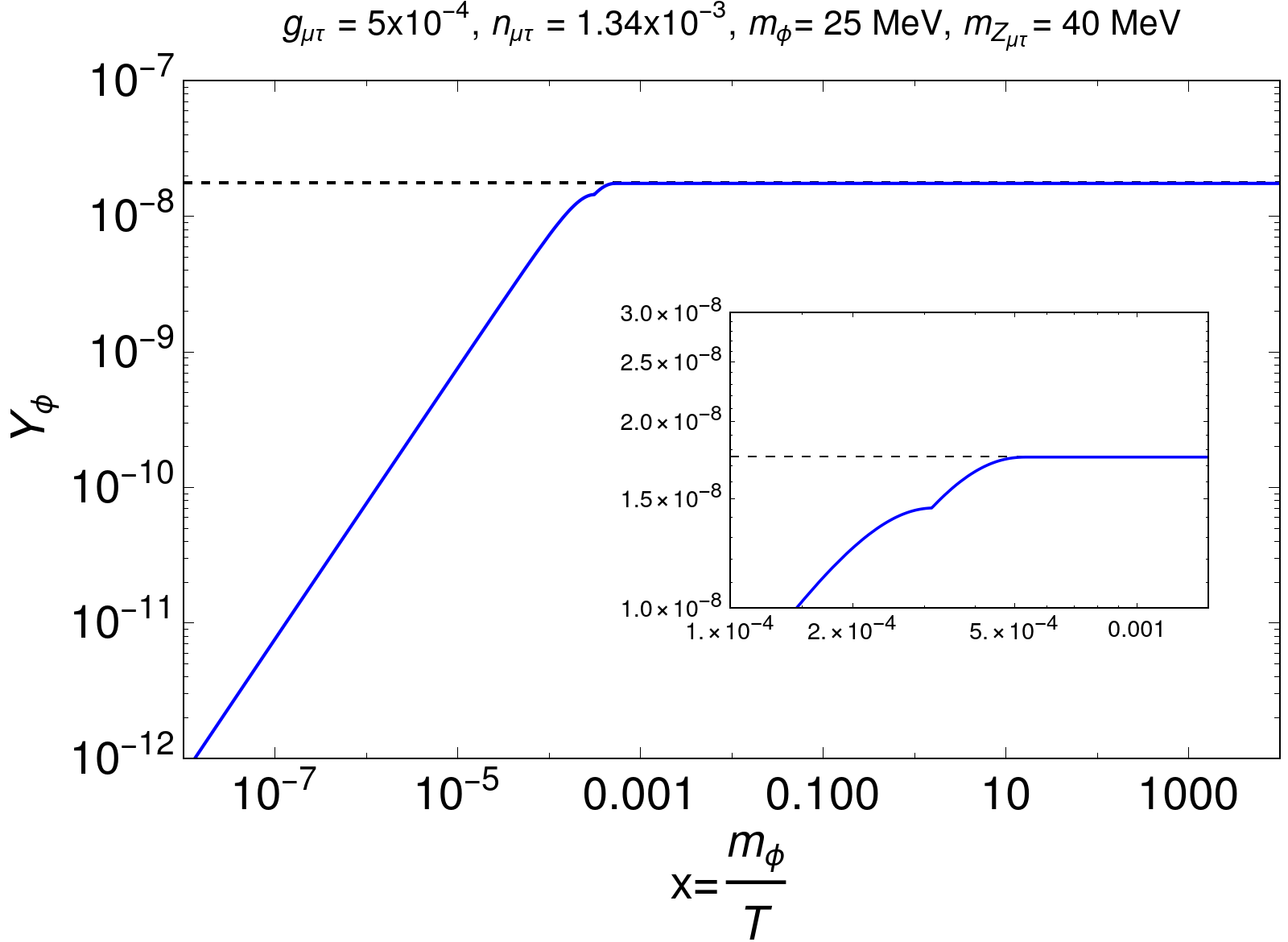}
		\caption{Evolution of the DM comoving number density {\color{blue}
		(blue line)} as a function of $x = \frac{m_\phi}{T}$ for $m_\phi$ = 25 MeV when DM production from scalar decay is negligible. The dashed horizontal line corresponds to the value of DM abundance ($Y_\phi$) for which the observed relic density is satisfied.}
		\label{yphi_wos_p025}
	\end{figure}

	Next, we plot the DM comoving number densities $Y_\phi(x)$ as a function of $x$ in Figs. \ref{yphi_wos_1},\ref{yphi_wos_p030} and \ref{yphi_wos_p025}.
		One key takeaway from this study is that thermal corrections to the mass of the decaying particle can open up new temperature thresholds not encountered in SFI.
		And the preceding discussion enables 
		an intuitive understanding of the DM yield as a function of temperature. First, we point out that in the absence of thermal corrections, 
		$Z_{\mu\tau}$ is either massless (for $T > v_{\mu\tau}$) or at best has 
		a 40 MeV mass ($T < v_{\mu\tau}$). That is, $m_{Z_{\mu\tau}} \leq 2 m_\phi$ in either case for all the $m_\phi$ values chosen. Therefore, the decay $Z_{\mu\tau} \to \phi \phi^\dagger$ remains kinematically closed, and no DM production must take place in the SFI picture. However, as detailed before, this decay mode kinematically opens up upon incorporating the thermal corrections, and DM production gets triggered and continues up to $x = x_1^{\text{cr}}$ for the DM masses permitting the first crossover. The decay threshold closes at $x=x_1^{cr}$, and DM production abruptly stops causing the DM yield to saturate at $Y_\phi(x_1^{\text{cr}})$ immediately after. For $m_\phi$ = 1 GeV, the $Z_{\mu\tau} \to \phi \phi^\dagger$ threshold does not reopen at any later point. And this explains the horizontal line to the right of $x^{\text{cr}}_1$ in Fig. \ref{yphi_wos_1}.  
	

	For $m_\phi$ = 30 MeV, DM production stops at the first crossover point, thereby causing the plateau immediately to the right of $x^{\text{cr}}_1 = 3.22 \times 10^{-4}$. For this BP, the $Z_{\mu\tau} \to \phi \phi^\dagger$ threshold reopens shortly after at the symmetry breaking point, and freeze-in production kicks in again. This reopening is what shows up as the kink in Fig. \ref{yphi_wos_p030} around $x = 3.75 \times 10^{-4}$. However, this second phase of DM production is rather short-lived and terminates permanently at the second crossover point. Hence a second horizontal region $x^{\text{cr}}_2 = 4.33 \times 10^{-4}$ onwards in Fig. \ref{yphi_wos_p030}.

	Lastly, we discuss the freeze-in production for $m_\phi$ = 25 MeV. In this case, DM production is unimpeded up to the second crossover point. Therefore, one expectedly finds a plateau starting at $x^{\text{cr}}_2 = 5.37 \times 10^{-4}$ in Fig.~\ref{yphi_wos_p025}. The symmetry breaking only leads to the minor cusp around the corresponding $x$-value, i.e., $x = 3.12 \times 10^{-4}$. 	
	
	For the DM $\phi$ being generated through the decay of $Z_{\mu\tau}$ only, 
		$Y_\phi(x_\infty)$
		and therefore $\Omega_\phi h^2 \propto n^2_{\mu\tau}$. Having displayed the various temperature thresholds in the Figs.~\ref{yphi_wos_1}, \ref{yphi_wos_p030} and \ref{yphi_wos_p025}, we subsequently tune $n_{\mu\tau}$ appropriately such that $Y_\phi(x_\infty)$ is in the requisite $\sim 10^{-11}$ ballpark. For example, we find the appropriate $n_{\mu\tau}$ = $1.47 \times 10^{-3},~1.46 \times 10^{-3}$ and $1.34 \times 10^{-3}$ for $m_\phi$ = 1 GeV, 30 MeV and 25 MeV respectively. We have checked that such $\mathcal{O}(10^{-3})$ values for $n_{\mu\tau}$, through kinetic mixing, lead to direct detection cross sections below the LZ limit for all the BPs.
		

	\subsection{$\l_{H\phi},\l_{S\phi} \sim n_{\mu\tau}g_{\mu\tau}$}
	In this section, we study the impact of the 
	$h_1,h_2 \to \phi \phi^\dagger$ decays on freeze-in production for the chosen $m_\phi$ values. It, therefore, becomes pertinent here to examine the thermal corrections to the masses of the decaying scalars. A crucial difference between $Z_{\mu\tau} \to \phi \phi^\dagger$ and $h_1,h_2 \to \phi \phi^\dagger$ in this model is that while the former can lead to DM production at very early epochs (or at a very high $T$) through thermal corrections, the latter is triggered primarily through spontaneous symmetry breaking. Given that we have $v_{\mu\tau}$ = 80 GeV in addition to $v$ = 246 GeV, one can treat $v_{\text{SB}} \sim$ 100 GeV
	as a common symmetry breaking scale, and therefore the scalar decays are activated for $T \leq v_{\text{SB}}$. We further take $M_{h_2}$ = 25 GeV, sin$\theta$ = 0.01 and $y_{e\mu}=y_{e\tau}$ = 0.5 consistently with the collider constraints and neutrino data. And the impact of thermal corrections on the scalar masses becomes subdominant for such a temperature range. We first quote below the thermally corrected masses for $h_{1,2}$ to test this impact. Neglecting the effect of a small $s_\theta$, and the couplings $\l_{H\phi}$ and $\l_{S\phi}$, one writes
	\besub
	\bea
	M^2_{h_1}(T) &\simeq& m^2_{h_1} + \frac{1}{12}\big(6\l_H + \l_{HS} + 3 y_t^2
	+ \frac{3}{4}(g^\prime)^2 + \frac{9}{4}g^2 \big)T^2, \\
	M^2_{h_2}(T) &\simeq& m^2_{h_2} + \frac{1}{12}(2 \l_{HS} + 4 \l_S + y^2_{e\mu}
	+ y^2_{e\tau} + 3 g^2_{\mu\tau})T^2.
	\eea
	\eesub
	This choice corresponds to $\l_H = 0.258,~\l_{S} = 0.098$ and $\l_{HS}=0.015$ from Eqs.(\ref{lH})-(\ref{lHS}). One then obtains $M_{h_1}(v_{\mu\tau})$ = 153.73 GeV and $M_{h_2}(v_{\mu\tau})$ = 33.42 GeV. Thus, for both $m_{h_1}$ = 125 GeV and $m_{h_2}$ = 25 GeV, the correction generated from the thermal loops is incremental. 
	\begin{figure}[htb!]
		\includegraphics[height=6.5cm,width=8cm]{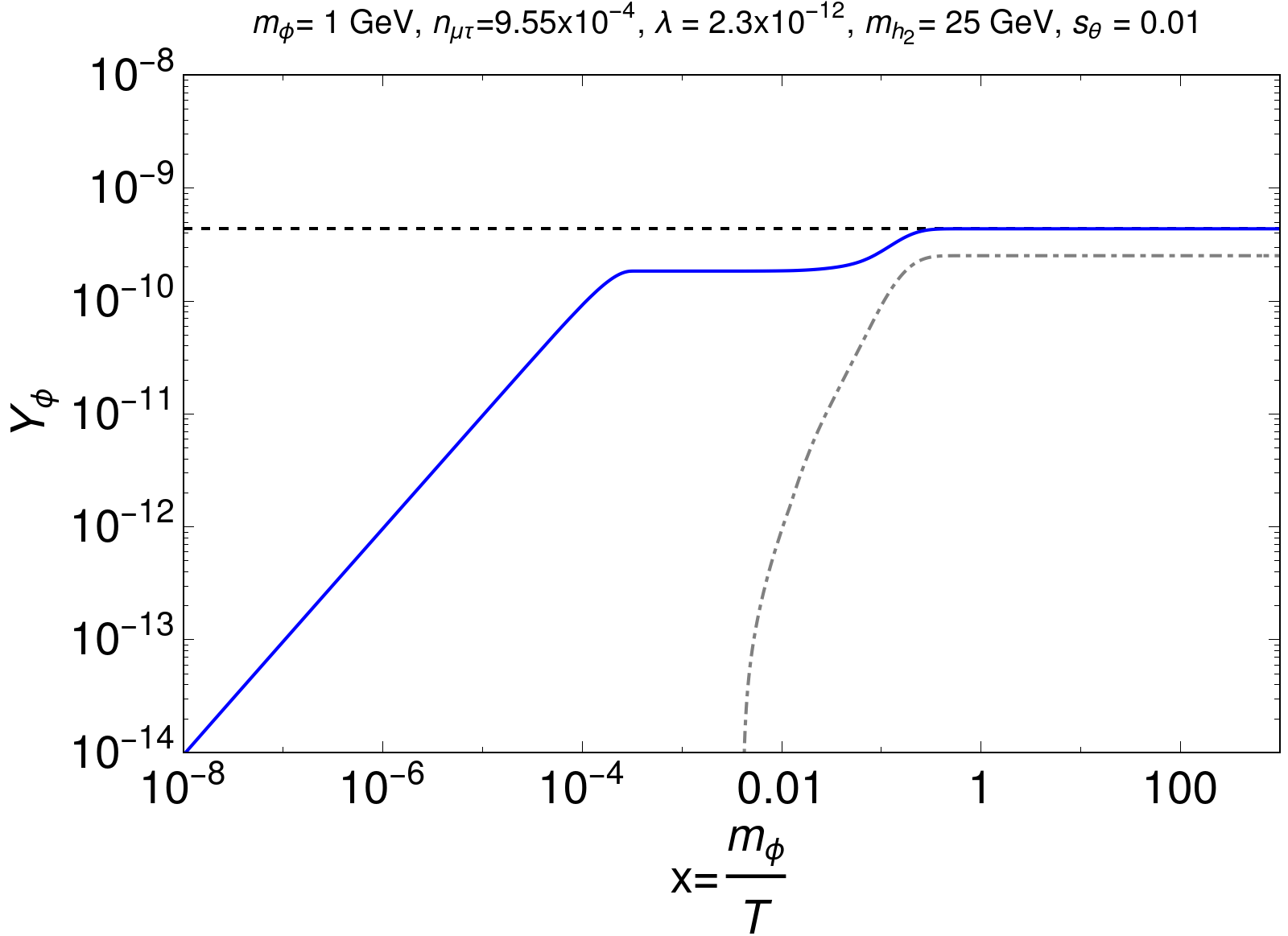}
		\caption{Evolution of the DM comoving number density {\color{blue}(blue line)} as a function of $x = \frac{m_\phi}{T}$ for $m_\phi$ = 1 GeV when scalar decays are not negligible. The dot-dashed line is the corresponding SFI curve. The dashed horizontal line corresponds to the value of DM abundance ($Y_\phi$) for which the observed relic density is satisfied.}
		\label{yphi_ws_1}
	\end{figure}
	Such choices for $m_\phi$ and $m_{h_2}$, therefore, imply that unlike $Z_{\mu\tau} \to \phi \phi^\dagger$, the $h_1,h_2 \to \phi \phi^\dagger$ decays can occur even in the absence of temperature effects. Therefore, DM production in the SFI picture is not completely ruled out for this subsection.

\begin{figure}[htb!]
\includegraphics[height=6.5cm,width=8cm]{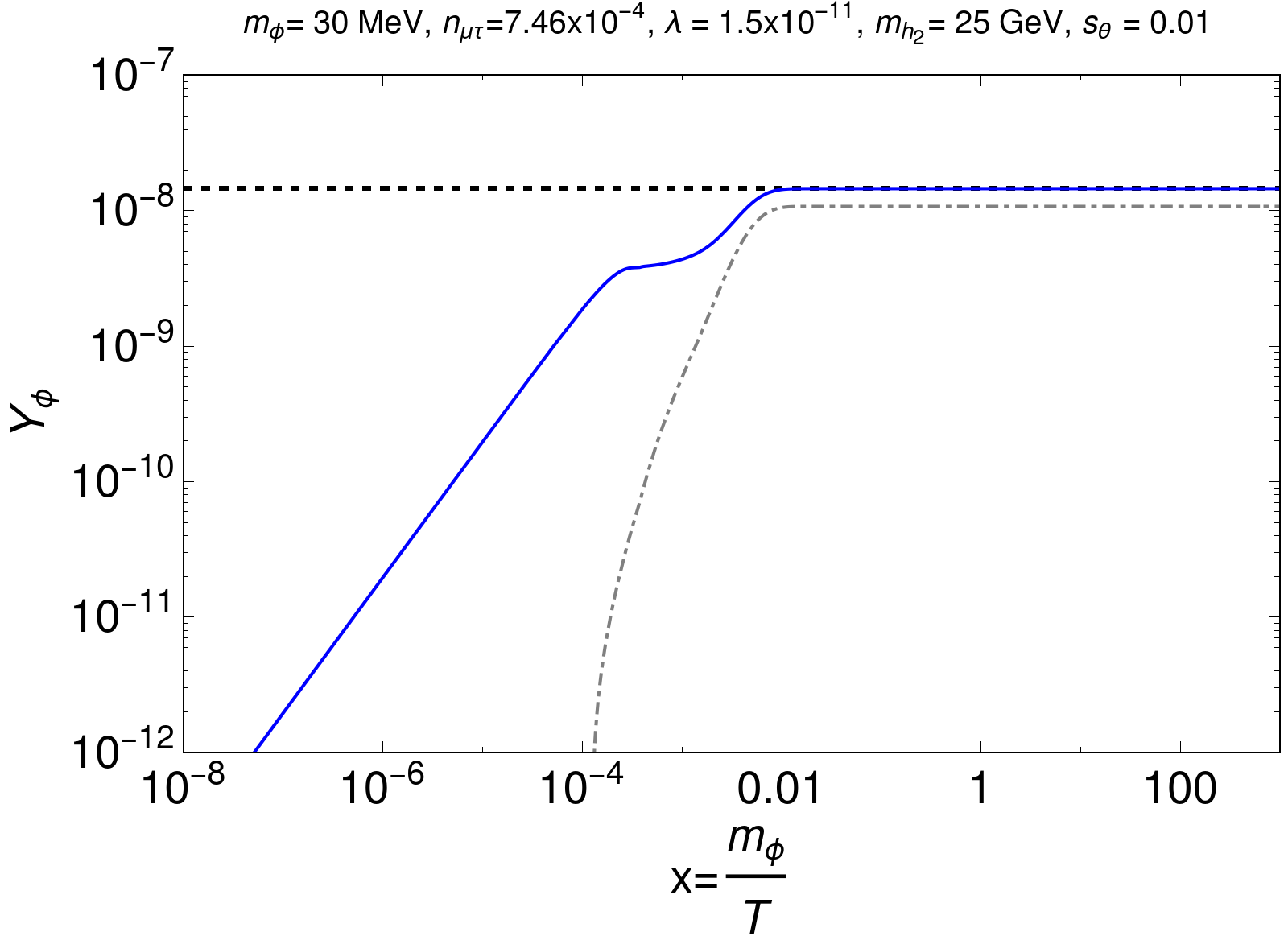}
		\caption{Evolution of the DM comoving number density {\color{blue}(blue line)} as a function of $x = \frac{m_\phi}{T}$ for $m_\phi$ = 30 MeV when scalar decays are not negligible. The dot-dashed line is the corresponding SFI curve. The dashed horizontal line corresponds to the value of DM abundance ($Y_\phi$) for which the observed relic density is satisfied.}
		\label{yphi_ws_p030}
	\end{figure}

\begin{figure}[htb!]
		\includegraphics[height=6.5cm,width=8cm]{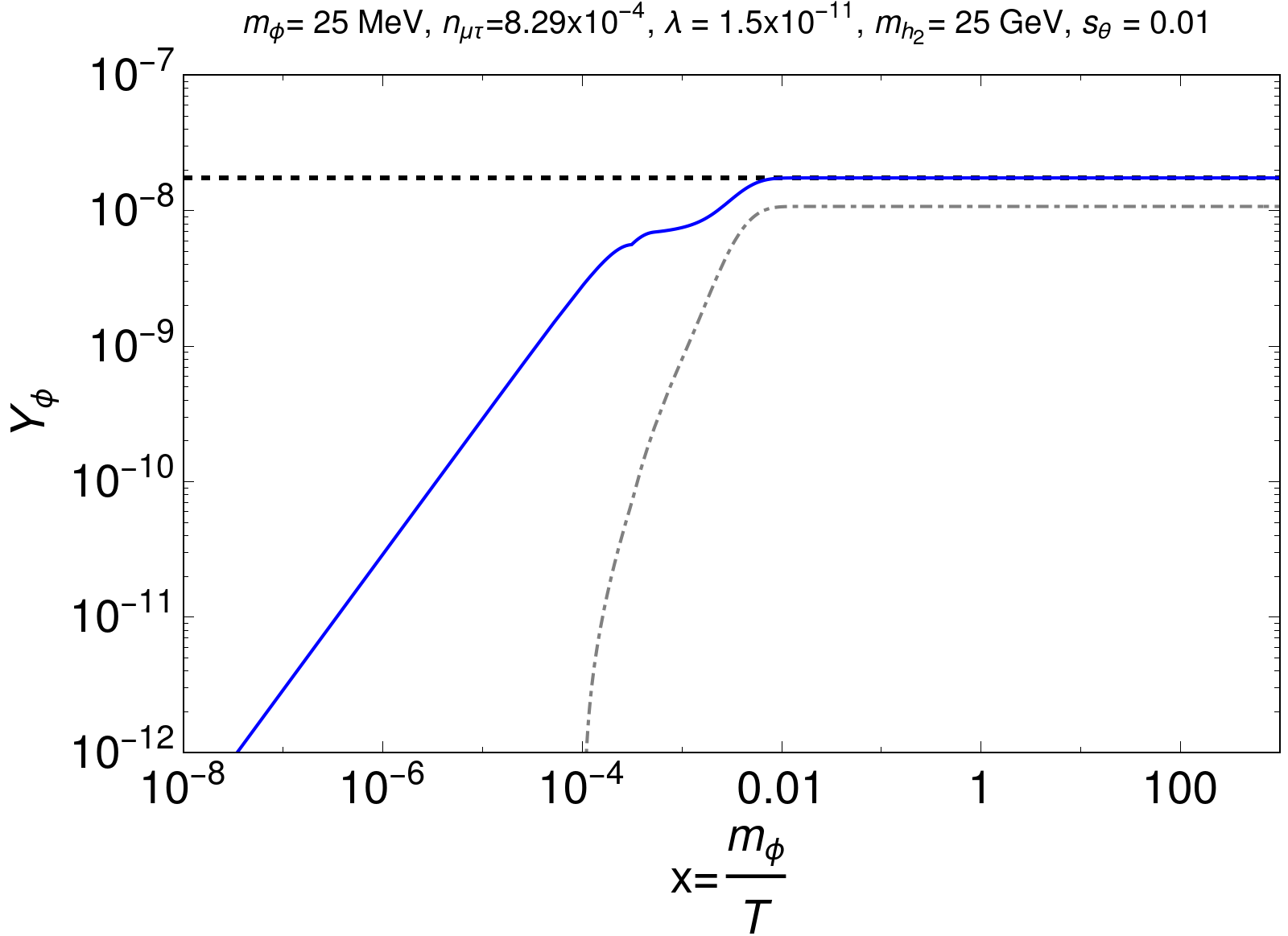}
		\caption{Evolution of the DM comoving number density {\color{blue}(blue line)} as a function of $x = \frac{m_\phi}{T}$ for $m_\phi$ = 25 MeV when scalar decays are not negligible. The dot-dashed line is the corresponding SFI curve. The dashed horizontal line corresponds to the value of DM abundance ($Y_\phi$) for which the observed relic density is satisfied.}
		\label{yphi_ws_p025}
	\end{figure}

Figs. \ref{yphi_ws_1}, \ref{yphi_ws_p030} and \ref{yphi_ws_p025} show the impact of scalar decays on the DM yield for $m_\phi$ = 1 GeV, 30 MeV, and 25 MeV respectively. For the chosen DM masses, both $M_{h_1}(T)$ and $M_{h_2}(T)$ remain greater than 2$m_\phi$ in the $T < v_{\mu\tau}$ range. In other words, the finite temperature corrections do not alter the original hierarchy in this case. We also take $\l_{H\phi} = \l_{S\phi} \equiv \l$ for simplicity. For $m_\phi$ = 1 GeV,  
		DM production from scalar decay kicks in around $x \simeq 0.01$, an epoch when the production from $Z_{\mu\tau}$ decay has already ceased long back. As can be seen in Fig. \ref{yphi_ws_1}, the scalar decays thus "lift" the horizontal DM yield curve, and the natural freeze-in saturation smoothly is attained around $x \sim \mathcal{O}(0.1)$. No new crossovers are introduced in the process. The cases of BP2 and BP3 are also not very different. For BP2 and BP3, DM production from $Z_{\mu\tau}$ decay stops at $x^{\text{cr}}_2 = 4.33 \times 10^{-4}$ and $5.37 \times 10^{-4}$ respectively. However $h_1,h_2 \to \phi \phi^\dagger$ get activated at a slightly earlier epoch, i.e., $x \simeq 3 \times 10^{-4}$. So, DM matter production never entirely ceases for BP2 and BP3. This is corroborated by Figs. \ref{yphi_ws_p030} and \ref{yphi_ws_p025}. Now that scalar decays too contribute to DM production, the parameters  $\l$ and $n_{\mu\tau}$ values need to be chosen carefully so as to obtain the required relic density. These parameters values for each BP can be read from the corresponding figure. For convenience, they are also listed in Table \ref{tab:quantify}. The same table also shows the percentage contribution to the observed relic through SFI for the three BPs. These numbers show that thermal correction to the mass of the decaying $Z_{\mu\tau}$ can indeed lead to sizeable increments in the relic. Lastly, we state for completeness that the values listed for $\l$ and $n_{\mu\tau}$ in Table \ref{tab:quantify} lead to direct detection cross sections well below the LZ limit.

\begin{table}[t]
	\small
	\centering
	\begin{tabular}{|l|c|c|c|c|}
		\hline
		Benchmark 	& $n_{\mu\tau}$ 	&  $\lambda$  &  SFI contribution & FFI contribution\\     \hline  \hline
		BP1  & $9.55\times 10^{-4}$ 		&  $2.3\times 10^{-12}$  			&	$58\%$		& $42\%$	\\     \hline
		BP2 & $7.46\times 10^{-4}$		&  $1.5\times 10^{-11}$ 		 	&$75\%$	& $25\%$	\\     \hline
		BP3 & $8.29\times 10^{-4}$		&  $1.5\times 10^{-11}$ &  $62\%$	&	$38\%$	\\     \hline
	\end{tabular}
	\caption{Percentage contribution of SFI and FFI in obtaining the observed DM relic abundance ($\Omega_{\phi}h^2=0.12$) for the benchmark points considered in the analysis. }
	\label{tab:quantify}
\end{table}

\begin{figure}[htb!]
		\includegraphics[scale=0.45]{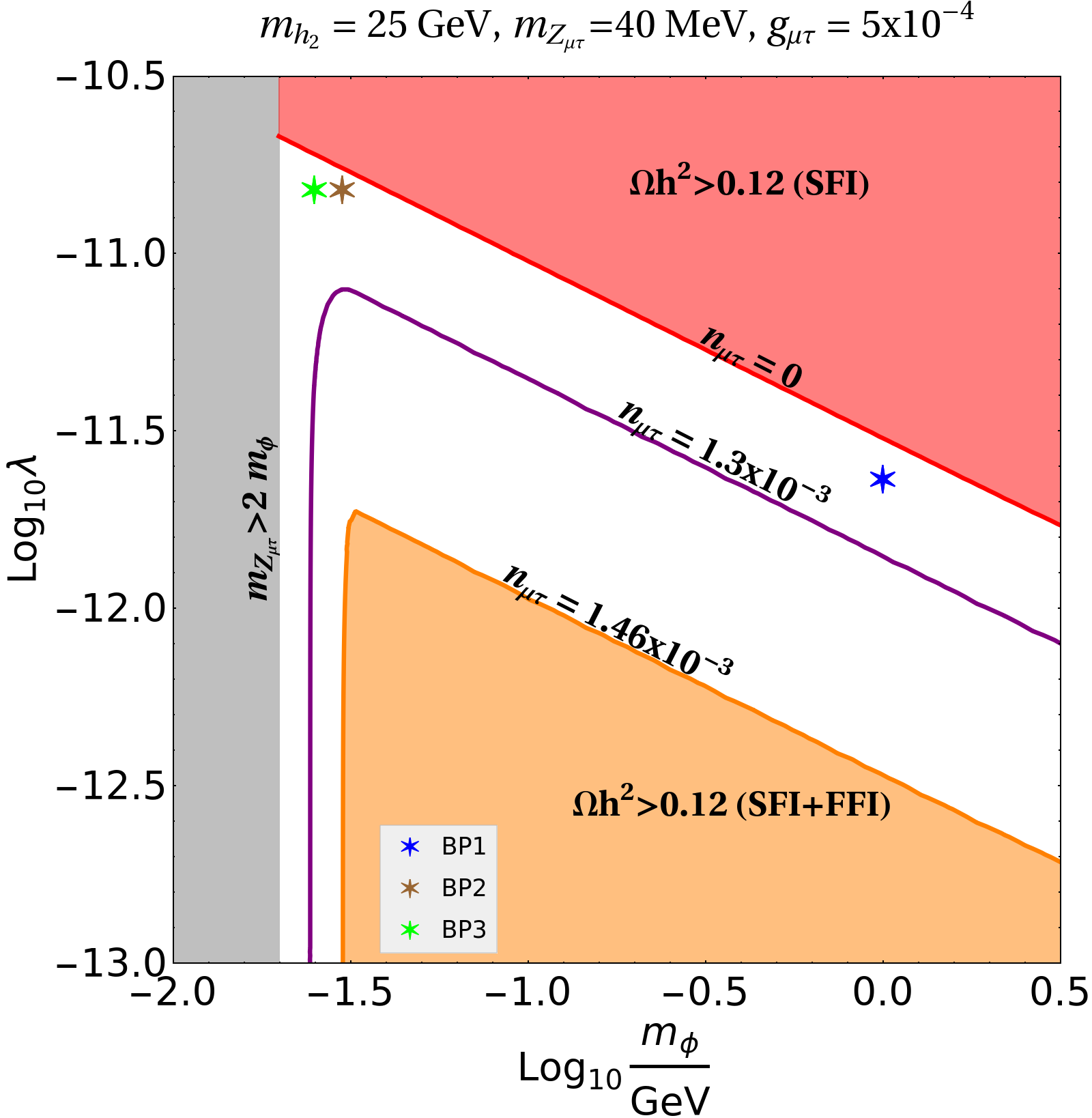}
		\caption{Allowed parameter space in $m_\phi-\lambda$ plane under different freeze-in schemes for fixed values of $m_{h_2},m_{Z_{\mu\tau}}~\text{and}~g_{\mu\tau}$. The grey region shows the parameter space where the DM can be produced via SFI from the decay of $Z_{\mu\tau}$ and hence is not the region of our interest. The red solid line signifies the parameter space for which the observed DM relic density can directly be obtained from SFI (scalars decay) and above which the DM relic always remains overabundant (pink region). The purple and orange solid lines correspond to a parameter space where the correct DM relic density is observed but as a result of an interplay between SFI and FFI for $n_{\mu\tau}=1.3\times10^{-3}$ and $n_{\mu\tau}=1.46\times10^{-3}$ respectively. The light orange region shows to parameter space where the relic again remains overabundant.  Finally, the different colored stars correspond to our BPs. }
		\label{par_constraint}
	\end{figure}

 Next, we use Fig.~\ref{par_constraint} to present a clear distinction between the SFI and FFI scenarios.  We present this plot in the 
$m_\phi-\lambda$ plane where we fix $m_{h_2}=25~\text{GeV},~m_{Z_{\mu\tau}}=40~\text{MeV},~\text{and}~g_{\mu\tau}=5\times10^{-4}$. To begin with, the grey region corresponds to $m_{Z_{\mu\tau}}>2m_\phi$. DM production can occur here through SFI and thus this region is not of our interest.
 The red solid line corresponds to $n_{\mu\tau}=0$ and here DM cannot be produced from the decay of $Z_{\mu\tau}$ even in presence of thermal corrections (through the forbidden channel) and the only source of DM production are the SFI channels (resulting from the scalar decays). As expected, one needs to decrease the value of quartic coupling $\lambda$ for an increasing DM mass in order to satisfy the observed DM relic density. The pink region above the red solid line shows the region of parameter space where the DM is always overproduced through the SFI channels. Whenever $n_{\mu\tau}$ is non-zero, DM production from the decay of $Z_{\mu\tau}$ through FFI also comes into the picture. Now, FFI together with the SFI helps the DM to satisfy the observed DM relic abundance. This behavior is also evident from Fig.~\ref{yphi_ws_1},~\ref{yphi_ws_p030} and \ref{yphi_ws_p025} which corresponds to our BP1, BP2 and BP3. These BPs are shown using different colored stars in Fig.~\ref{par_constraint}. It is interesting to point out that increasing $n_{\mu\tau}$ gradually increases the contribution of $Z_{\mu\tau}$ decay towards the DM abundance and hence the contributions from the scalar decay (SFI) has to be appropriately tuned in order to satisfy the correct DM relic abundance. This behavior is depicted by the purple curve where we set $n_{\mu\tau}=1.3\times10^{-3}$. We notice that this curve remains independent of $\lambda$ for any $\lambda\leq 8\times10^{-12}$ for DM mass 24 MeV. This is because, for this combination of $n_{
\mu\tau}$ and $m_{\phi}$, the DM production from the $Z_{\mu\tau}$ decay can itself give $100\%$ contribution towards the DM relic density (production of DM from the scalar decay remains subdominant). Increasing $\lambda$ above $ 8\times10^{-12}$ can results in an overproduction of the DM as the contribution from scalar decay (SFI channels) towards DM production also becomes significant. Once the peak is reached the curve starts to fall with the increasing DM mass. This is expected as in this regime, both FFI and SFI channels contribute towards the DM production to satisfy the correct relic abundance. Increasing $n_{\mu\tau}$ further leads to an increase in the DM relic abundance for a fixed DM mass and hence a smaller $\lambda$ is required in order to satisfy the observed DM relic abundance. This is clear from the orange curve where $n_{\mu\tau}=1.46\times10^{-3}$. The trend observed for non-zero $n_{\mu\tau}$ in $m_\phi-\lambda$ plane also suggests that there exists an upper bound on the combination of $n_{\mu\tau}$ and $m_{\phi}$ (which corresponds to a vertical line) that can give the correct DM relic abundance with a dominant contribution coming from the decay of $Z_{\mu\tau}$ and going beyond it always results in an overabundance of DM (light orange region). For this particular choice of fixed parameters, we found the combination to be $n_{\mu\tau}=1.46\times10^{-3}$ and $m_\phi=32$ MeV (orange curve). Next, we would also like to point out the non-trivial shift observed in the DM mass when $n_{\mu\tau}$ is increased from a smaller to a larger value (shift in the vertical line). Here, we notice that a larger $n_{\mu\tau}$ requires a larger DM mass to satisfy the observed DM relic density. The reason for this behavior is the following, for a larger DM mass the first crossover of $M_{Z_{\mu\tau}}(T)$ and $2m_\phi$ takes place at an earlier epoch in comparison to what happens for a smaller DM mass. This early crossover results in a smaller DM abundance when produced from the decay of $Z_{\mu\tau}$. Hence demanding the observed relic density accordingly requires a larger $n_{\mu\tau}$.

	Finally, we will also like to comment briefly on the possibility of having a large $v_{\mu\tau}$ with the same choice of $g_{\mu\tau}$ fixed at $5\times10^{-4}$ as discussed above. A large $v_{\mu\tau}$ results in a heavier $Z_{\mu\tau}$ (vertical part of the orange curve). Although this scenario does not remain consistent with the recent $g-2$ data, as can also be seen from Fig.~\ref{f:const}, it still is interesting in terms of DM phenomenology. If the mass hierarchy among the $Z_{\mu\tau}$, the scalar responsible for breaking $U(1)_{L_\mu-L_\tau}$ symmetry, and the DM are appropriately set the scenario can result in the forbidden production of the DM from the decay of this scalar at a very early epoch through thermal corrections. The production of DM from $Z_{\mu\tau}$ decay will always proceed through the SFI even with the thermal mass of $Z_{\mu\tau}$ is taken into account if the DM mass is smaller than half of the $Z_{\mu\tau}$ mass. Discussing this scenario in detail is beyond the scope of the present work, and we wish to take it as a future project.
      \section{Summary and Conclusion}
	\label{conclusion}
	
	Although the studies of FIMP dark matter in a minimally extended $U(1)_{L_\mu-L_\tau}$ model already exists in the literature, the role of thermal corrections in this setup has never been examined before. In this work, we show that incorporating a thermal mass for the gauge boson $Z_{\mu\tau}$ opens up new temperature thresholds that are not encountered in SFI. For simplicity, we only consider the production of the DM through the decay of the gauge boson of $U(1)_{L_\mu-L_\tau}$ symmetry and two scalars. All the above-mentioned particles remain in equilibrium with the SM bath but couple feebly to the DM. The DM mass does not receive thermal corrections on account of the fact that a FIMP does not equilibrate with the thermal bath at any point. However, the masses of the decaying particles receive such corrections at high temperatures due to their interactions with the thermal bath. 
	
	For a better understanding of the role of different thermal masses, we divide our study into two different scenarios: (A) $\l_{H\phi},\l_{S\phi}<< n_{\mu\tau}g_{\mu\tau}$ and (B) $\l_{H\phi},\l_{S\phi}\sim n_{\mu\tau}g_{\mu\tau}$.  In the first scenario, the DM is dominantly produced by the decay of $Z_{\mu\tau}$ whereas the second scenario entails the production of DM from the decay of $Z_{\mu\tau}$ as well as the other two scalars. An exciting feature of the first scenario is the existence of two crossovers where the condition $M_{Z_{\mu\tau}}(T)>2m_{\phi}$ is satisfied. While the DM production always proceeds via a channel that remains kinematically forbidden in the SFI scenario before the first crossover, and the production of the DM after the second crossover might or might not happen via the forbidden channel. In the second scenario, the impact of the scalars decaying into the DM is also observed on top of its production from the decay of $Z_{\mu\tau}$. Here, the production from the scalar decay is triggered only after the spontaneous symmetry breaking. Finally, we also comment on the possibility of having a larger $U(1)_{L_\mu-L_\tau}$ breaking scale, which we wish to take as a future endeavor. 
	
	The involvement of the feeble interactions of DM with the SM particles in the freeze-in scenario makes the model exceedingly challenging to observe experimentally. With the WIMP DM parameter space almost getting ruled out from the present experiments, the FIMP-type DM has emerged as a new alternative. Hence, providing a detection prospect of such DM is always a compelling task. Keeping this in mind, we focused on a DM parameter of the present setup that remained consistent with the DM relic density but at the same time also provided a solution to the muon $(g-2)$ anomaly. This, in turn, also increases the predictability of the present setup.

	\acknowledgments
	This work is supported by the Physical Research Laboratory (PRL), Department of Space, Government of India. Computational work was performed using the HPC resources (Vikram-100 HPC) and the TDP project at PRL. NC acknowledges support from DST, India, under grant number IFA19- PH237 (INSPIRE Faculty Award). R.R. also acknowledges the National Research Foundation of Korea (NRF) grant funded by the Korean government (NRF-2020R1C1C1012452). 

	\bibliographystyle{JHEP}
	\bibliography{ref}

\providecommand{\href}[2]{#2}\begingroup\raggedright\begin{thebibliography}{10}

\bibitem{Sofue:2000jx}
Y.~Sofue and V.~Rubin, \emph{{Rotation curves of spiral galaxies}},
  \href{http://dx.doi.org/10.1146/annurev.astro.39.1.137}{\emph{Ann. Rev.
  Astron. Astrophys.} {\bf 39} (2001) 137--174},
  [\href{http://arxiv.org/abs/astro-ph/0010594}{{\tt astro-ph/0010594}}].

\bibitem{Clowe:2006eq}
D.~Clowe, M.~Bradac, A.~H. Gonzalez, M.~Markevitch, S.~W. Randall, C.~Jones
  et~al., \emph{{A direct empirical proof of the existence of dark matter}},
  \href{http://dx.doi.org/10.1086/508162}{\emph{Astrophys. J. Lett.} {\bf 648}
  (2006) L109--L113}, [\href{http://arxiv.org/abs/astro-ph/0608407}{{\tt
  astro-ph/0608407}}].

\bibitem{Planck:2018vyg}
{\scshape Planck} collaboration, N.~Aghanim et~al., \emph{{Planck 2018 results.
  VI. Cosmological parameters}},
  \href{http://dx.doi.org/10.1051/0004-6361/201833910}{\emph{Astron.
  Astrophys.} {\bf 641} (2020) A6},
  [\href{http://arxiv.org/abs/1807.06209}{{\tt 1807.06209}}].

\bibitem{Hinshaw_2013}
G.~Hinshaw, D.~Larson, E.~Komatsu, D.~N. Spergel, C.~L. Bennett, J.~Dunkley
  et~al., \emph{Nine-year wilkinson microwave anisotropy probe ( wmap )
  observations: Cosmological parameter results},
  \href{http://dx.doi.org/10.1088/0067-0049/208/2/19}{\emph{The Astrophysical
  Journal Supplement Series} {\bf 208} (Sep, 2013) 19}.

\bibitem{Konar:2020wvl}
P.~Konar, A.~Mukherjee, A.~K. Saha and S.~Show, \emph{{Linking pseudo-Dirac
  dark matter to radiative neutrino masses in a singlet-doublet scenario}},
  \href{http://dx.doi.org/10.1103/PhysRevD.102.015024}{\emph{Phys. Rev. D} {\bf
  102} (2020) 015024}, [\href{http://arxiv.org/abs/2001.11325}{{\tt
  2001.11325}}].

\bibitem{Konar:2020vuu}
P.~Konar, A.~Mukherjee, A.~K. Saha and S.~Show, \emph{{A dark clue to seesaw
  and leptogenesis in a pseudo-Dirac singlet doublet scenario with
  (non)standard cosmology}},
  \href{http://dx.doi.org/10.1007/JHEP03(2021)044}{\emph{JHEP} {\bf 03} (2021)
  044}, [\href{http://arxiv.org/abs/2007.15608}{{\tt 2007.15608}}].

\bibitem{Barman:2021qds}
B.~Barman, N.~Bernal, A.~Das and R.~Roshan, \emph{{Non-minimally Coupled Vector
  Boson Dark Matter}},  \href{http://arxiv.org/abs/2108.13447}{{\tt
  2108.13447}}.

\bibitem{Chakrabarty:2021kmr}
N.~Chakrabarty, R.~Roshan and A.~Sil, \emph{{Two Component Doublet-Triplet
  Scalar Dark Matter stabilising the Electroweak vacuum}},
  \href{http://arxiv.org/abs/2102.06032}{{\tt 2102.06032}}.

\bibitem{DuttaBanik:2020jrj}
A.~Dutta~Banik, R.~Roshan and A.~Sil, \emph{{Two component singlet-triplet
  scalar dark matter and electroweak vacuum stability}},
  \href{http://dx.doi.org/10.1103/PhysRevD.103.075001}{\emph{Phys. Rev. D} {\bf
  103} (2021) 075001}, [\href{http://arxiv.org/abs/2009.01262}{{\tt
  2009.01262}}].

\bibitem{Borah:2020nsz}
D.~Borah, R.~Roshan and A.~Sil, \emph{{Sub-TeV singlet scalar dark matter and
  electroweak vacuum stability with vectorlike fermions}},
  \href{http://dx.doi.org/10.1103/PhysRevD.102.075034}{\emph{Phys. Rev. D} {\bf
  102} (2020) 075034}, [\href{http://arxiv.org/abs/2007.14904}{{\tt
  2007.14904}}].

\bibitem{Bhattacharya:2019tqq}
S.~Bhattacharya, N.~Chakrabarty, R.~Roshan and A.~Sil, \emph{{Multicomponent
  dark matter in extended $U(1)_{B-L}$: neutrino mass and high scale
  validity}},
  \href{http://dx.doi.org/10.1088/1475-7516/2020/04/013}{\emph{JCAP} {\bf 04}
  (2020) 013}, [\href{http://arxiv.org/abs/1910.00612}{{\tt 1910.00612}}].

\bibitem{Borah:2019aeq}
D.~Borah, R.~Roshan and A.~Sil, \emph{{Minimal two-component scalar doublet
  dark matter with radiative neutrino mass}},
  \href{http://dx.doi.org/10.1103/PhysRevD.100.055027}{\emph{Phys. Rev. D} {\bf
  100} (2019) 055027}, [\href{http://arxiv.org/abs/1904.04837}{{\tt
  1904.04837}}].

\bibitem{Baek:2008nz}
S.~Baek and P.~Ko, \emph{{Phenomenology of U(1)(L(mu)-L(tau)) charged dark
  matter at PAMELA and colliders}},
  \href{http://dx.doi.org/10.1088/1475-7516/2009/10/011}{\emph{JCAP} {\bf 10}
  (2009) 011}, [\href{http://arxiv.org/abs/0811.1646}{{\tt 0811.1646}}].

\bibitem{Patra:2016shz}
S.~Patra, S.~Rao, N.~Sahoo and N.~Sahu, \emph{{Gauged $U(1)_{L_\mu - L_\tau}$
  model in light of muon $g-2$ anomaly, neutrino mass and dark matter
  phenomenology}},
  \href{http://dx.doi.org/10.1016/j.nuclphysb.2017.02.010}{\emph{Nucl. Phys. B}
  {\bf 917} (2017) 317--336}, [\href{http://arxiv.org/abs/1607.04046}{{\tt
  1607.04046}}].

\bibitem{Biswas:2016yan}
A.~Biswas, S.~Choubey and S.~Khan, \emph{{Neutrino Mass, Dark Matter and
  Anomalous Magnetic Moment of Muon in a $U(1)_{L_{\mu}-L_{\tau}}$ Model}},
  \href{http://dx.doi.org/10.1007/JHEP09(2016)147}{\emph{JHEP} {\bf 09} (2016)
  147}, [\href{http://arxiv.org/abs/1608.04194}{{\tt 1608.04194}}].

\bibitem{LUX:2015abn}
{\scshape LUX} collaboration, D.~S. Akerib et~al., \emph{{Improved Limits on
  Scattering of Weakly Interacting Massive Particles from Reanalysis of 2013
  LUX Data}},
  \href{http://dx.doi.org/10.1103/PhysRevLett.116.161301}{\emph{Phys. Rev.
  Lett.} {\bf 116} (2016) 161301}, [\href{http://arxiv.org/abs/1512.03506}{{\tt
  1512.03506}}].

\bibitem{Akerib:2016vxi}
{\scshape LUX} collaboration, D.~S. Akerib et~al., \emph{{Results from a search
  for dark matter in the complete LUX exposure}},
  \href{http://dx.doi.org/10.1103/PhysRevLett.118.021303}{\emph{Phys. Rev.
  Lett.} {\bf 118} (2017) 021303}, [\href{http://arxiv.org/abs/1608.07648}{{\tt
  1608.07648}}].

\bibitem{Zhang:2018xdp}
{\scshape PandaX} collaboration, H.~Zhang et~al., \emph{{Dark matter direct
  search sensitivity of the PandaX-4T experiment}},
  \href{http://dx.doi.org/10.1007/s11433-018-9259-0}{\emph{Sci. China Phys.
  Mech. Astron.} {\bf 62} (2019) 31011},
  [\href{http://arxiv.org/abs/1806.02229}{{\tt 1806.02229}}].

\bibitem{Aprile:2018dbl}
{\scshape XENON} collaboration, E.~Aprile et~al., \emph{{Dark Matter Search
  Results from a One Ton-Year Exposure of XENON1T}},
  \href{http://dx.doi.org/10.1103/PhysRevLett.121.111302}{\emph{Phys. Rev.
  Lett.} {\bf 121} (2018) 111302}, [\href{http://arxiv.org/abs/1805.12562}{{\tt
  1805.12562}}].

\bibitem{LZ:2022ufs}
{\scshape LZ} collaboration, J.~Aalbers et~al., \emph{{First Dark Matter Search
  Results from the LUX-ZEPLIN (LZ) Experiment}},
  \href{http://arxiv.org/abs/2207.03764}{{\tt 2207.03764}}.

\bibitem{Hall:2009bx}
L.~J. Hall, K.~Jedamzik, J.~March-Russell and S.~M. West, \emph{{Freeze-In
  Production of FIMP Dark Matter}},
  \href{http://dx.doi.org/10.1007/JHEP03(2010)080}{\emph{JHEP} {\bf 03} (2010)
  080}, [\href{http://arxiv.org/abs/0911.1120}{{\tt 0911.1120}}].

\bibitem{Elahi:2014fsa}
F.~Elahi, C.~Kolda and J.~Unwin, \emph{{UltraViolet Freeze-in}},
  \href{http://dx.doi.org/10.1007/JHEP03(2015)048}{\emph{JHEP} {\bf 03} (2015)
  048}, [\href{http://arxiv.org/abs/1410.6157}{{\tt 1410.6157}}].

\bibitem{Bernal:2017kxu}
N.~Bernal, M.~Heikinheimo, T.~Tenkanen, K.~Tuominen and V.~Vaskonen, \emph{{The
  Dawn of FIMP Dark Matter: A Review of Models and Constraints}},
  \href{http://dx.doi.org/10.1142/S0217751X1730023X}{\emph{Int. J. Mod. Phys.
  A} {\bf 32} (2017) 1730023}, [\href{http://arxiv.org/abs/1706.07442}{{\tt
  1706.07442}}].

\bibitem{Darme:2019wpd}
L.~Darm\'e, A.~Hryczuk, D.~Karamitros and L.~Roszkowski, \emph{{Forbidden
  frozen-in dark matter}},
  \href{http://dx.doi.org/10.1007/JHEP11(2019)159}{\emph{JHEP} {\bf 11} (2019)
  159}, [\href{http://arxiv.org/abs/1908.05685}{{\tt 1908.05685}}].

\bibitem{Barman:2020plp}
B.~Barman, D.~Borah and R.~Roshan, \emph{{Effective Theory of Freeze-in Dark
  Matter}}, \href{http://dx.doi.org/10.1088/1475-7516/2020/11/021}{\emph{JCAP}
  {\bf 11} (2020) 021}, [\href{http://arxiv.org/abs/2007.08768}{{\tt
  2007.08768}}].

\bibitem{Barman:2021tgt}
B.~Barman, D.~Borah and R.~Roshan, \emph{{Nonthermal leptogenesis and UV
  freeze-in of dark matter: Impact of inflationary reheating}},
  \href{http://dx.doi.org/10.1103/PhysRevD.104.035022}{\emph{Phys. Rev. D} {\bf
  104} (2021) 035022}, [\href{http://arxiv.org/abs/2103.01675}{{\tt
  2103.01675}}].

\bibitem{Datta:2021elq}
A.~Datta, R.~Roshan and A.~Sil, \emph{{Imprint of the seesaw mechanism on
  feebly interacting dark matter and the baryon asymmetry}},
  \href{http://arxiv.org/abs/2104.02030}{{\tt 2104.02030}}.

\bibitem{Bhattacharya:2021jli}
S.~Bhattacharya, R.~Roshan, A.~Sil and D.~Vatsyayan, \emph{{Symmetry origin of
  Baryon Asymmetry, Dark Matter and Neutrino Mass}},
  \href{http://arxiv.org/abs/2105.06189}{{\tt 2105.06189}}.

\bibitem{Konar:2021oye}
P.~Konar, R.~Roshan and S.~Show, \emph{{Freeze-in dark matter through forbidden
  channel in U(1)\_B-L}},
  \href{http://dx.doi.org/10.1088/1475-7516/2022/03/021}{\emph{JCAP} {\bf 03}
  (2022) 021}, [\href{http://arxiv.org/abs/2110.14411}{{\tt 2110.14411}}].

\bibitem{Ghosh:2021wrk}
P.~Ghosh, P.~Konar, A.~K. Saha and S.~Show, \emph{{Self-interacting freeze-in
  dark matter in a singlet doublet scenario}},
  \href{http://arxiv.org/abs/2112.09057}{{\tt 2112.09057}}.

\bibitem{PhysRevD.45.2933}
M.~E. Carrington, \emph{Effective potential at finite temperature in the
  standard model},
  \href{http://dx.doi.org/10.1103/PhysRevD.45.2933}{\emph{Phys. Rev. D} {\bf
  45} (Apr, 1992) 2933--2944}.

\bibitem{Giudice:2003jh}
G.~F. Giudice, A.~Notari, M.~Raidal, A.~Riotto and A.~Strumia, \emph{{Towards a
  complete theory of thermal leptogenesis in the SM and MSSM}},
  \href{http://dx.doi.org/10.1016/j.nuclphysb.2004.02.019}{\emph{Nucl. Phys. B}
  {\bf 685} (2004) 89--149}, [\href{http://arxiv.org/abs/hep-ph/0310123}{{\tt
  hep-ph/0310123}}].

\bibitem{Laine:2016hma}
M.~Laine and A.~Vuorinen, \emph{{Basics of Thermal Field Theory}}, vol.~925.
\newblock Springer, 2016,
  \href{http://dx.doi.org/10.1007/978-3-319-31933-9}{10.1007/978-3-319-31933-9}.

\bibitem{Biondini:2020ric}
S.~Biondini and J.~Ghiglieri, \emph{{Freeze-in produced dark matter in the
  ultra-relativistic regime}},
  \href{http://dx.doi.org/10.1088/1475-7516/2021/03/075}{\emph{JCAP} {\bf 03}
  (2021) 075}, [\href{http://arxiv.org/abs/2012.09083}{{\tt 2012.09083}}].

\bibitem{Baker:2017zwx}
M.~J. Baker, M.~Breitbach, J.~Kopp and L.~Mittnacht, \emph{{Dynamic Freeze-In:
  Impact of Thermal Masses and Cosmological Phase Transitions on Dark Matter
  Production}}, \href{http://dx.doi.org/10.1007/JHEP03(2018)114}{\emph{JHEP}
  {\bf 03} (2018) 114}, [\href{http://arxiv.org/abs/1712.03962}{{\tt
  1712.03962}}].

\bibitem{Ramazanov:2021eya}
S.~Ramazanov, E.~Babichev, D.~Gorbunov and A.~Vikman, \emph{{Beyond freeze-in:
  Dark matter via inverse phase transition and gravitational wave signal}},
  \href{http://dx.doi.org/10.1103/PhysRevD.105.063530}{\emph{Phys. Rev. D} {\bf
  105} (2022) 063530}, [\href{http://arxiv.org/abs/2104.13722}{{\tt
  2104.13722}}].

\bibitem{He:1990pn}
X.~G. He, G.~C. Joshi, H.~Lew and R.~R. Volkas, \emph{{NEW Z-prime
  PHENOMENOLOGY}}, \href{http://dx.doi.org/10.1103/PhysRevD.43.R22}{\emph{Phys.
  Rev. D} {\bf 43} (1991) 22--24}.

\bibitem{He:1991qd}
X.-G. He, G.~C. Joshi, H.~Lew and R.~R. Volkas, \emph{{Simplest Z-prime
  model}}, \href{http://dx.doi.org/10.1103/PhysRevD.44.2118}{\emph{Phys. Rev.
  D} {\bf 44} (1991) 2118--2132}.

\bibitem{Biswas:2016yjr}
A.~Biswas, S.~Choubey and S.~Khan, \emph{{FIMP and Muon ($g-2$) in a
  U$(1)_{L_{\mu}-L_{\tau}}$ Model}},
  \href{http://dx.doi.org/10.1007/JHEP02(2017)123}{\emph{JHEP} {\bf 02} (2017)
  123}, [\href{http://arxiv.org/abs/1612.03067}{{\tt 1612.03067}}].

\bibitem{Banerjee:2018eaf}
H.~Banerjee, P.~Byakti and S.~Roy, \emph{{Supersymmetric gauged
  U(1)$_{L_{\mu}-L_{\tau}}$ model for neutrinos and the muon $(g-2)$ anomaly}},
  \href{http://dx.doi.org/10.1103/PhysRevD.98.075022}{\emph{Phys. Rev. D} {\bf
  98} (2018) 075022}, [\href{http://arxiv.org/abs/1805.04415}{{\tt
  1805.04415}}].

\bibitem{Chun:2018ibr}
E.~J. Chun, A.~Das, J.~Kim and J.~Kim, \emph{{Searching for flavored gauge
  bosons}}, \href{http://dx.doi.org/10.1007/JHEP02(2019)093}{\emph{JHEP} {\bf
  02} (2019) 093}, [\href{http://arxiv.org/abs/1811.04320}{{\tt 1811.04320}}].

\bibitem{Costa:2022oaa}
F.~Costa, S.~Khan and J.~Kim, \emph{{A two-component dark matter model and its
  associated gravitational waves}},
  \href{http://dx.doi.org/10.1007/JHEP06(2022)026}{\emph{JHEP} {\bf 06} (2022)
  026}, [\href{http://arxiv.org/abs/2202.13126}{{\tt 2202.13126}}].

\bibitem{Okada:2021nwo}
H.~Okada, Y.~Orikasa and Y.~Shoji, \emph{{Radiative dark matter and neutrino
  masses from an alternative U(1) B-L gauge symmetry}},
  \href{http://dx.doi.org/10.1088/1475-7516/2021/07/006}{\emph{JCAP} {\bf 07}
  (2021) 006}, [\href{http://arxiv.org/abs/2102.10944}{{\tt 2102.10944}}].

\bibitem{Hapitas:2021ilr}
T.~Hapitas, D.~Tuckler and Y.~Zhang, \emph{{General kinetic mixing in gauged
  U(1)L\ensuremath{\mu}-L\ensuremath{\tau} model for muon g-2 and dark
  matter}}, \href{http://dx.doi.org/10.1103/PhysRevD.105.016014}{\emph{Phys.
  Rev. D} {\bf 105} (2022) 016014},
  [\href{http://arxiv.org/abs/2108.12440}{{\tt 2108.12440}}].

\bibitem{CHARM-II:1990dvf}
{\scshape CHARM-II} collaboration, D.~Geiregat et~al., \emph{{First observation
  of neutrino trident production}},
  \href{http://dx.doi.org/10.1016/0370-2693(90)90146-W}{\emph{Phys. Lett. B}
  {\bf 245} (1990) 271--275}.

\bibitem{CCFR:1991lpl}
{\scshape CCFR} collaboration, S.~R. Mishra et~al., \emph{{Neutrino tridents
  and W Z interference}},
  \href{http://dx.doi.org/10.1103/PhysRevLett.66.3117}{\emph{Phys. Rev. Lett.}
  {\bf 66} (1991) 3117--3120}.

\bibitem{Altmannshofer:2014pba}
W.~Altmannshofer, S.~Gori, M.~Pospelov and I.~Yavin, \emph{{Neutrino Trident
  Production: A Powerful Probe of New Physics with Neutrino Beams}},
  \href{http://dx.doi.org/10.1103/PhysRevLett.113.091801}{\emph{Phys. Rev.
  Lett.} {\bf 113} (2014) 091801}, [\href{http://arxiv.org/abs/1406.2332}{{\tt
  1406.2332}}].

\bibitem{doi:10.1126/science.aao0990}
D.~Akimov, J.~B. Albert, P.~An, C.~Awe, P.~S. Barbeau, B.~Becker et~al.,
  \emph{Observation of coherent elastic neutrino-nucleus scattering},
  \href{http://dx.doi.org/10.1126/science.aao0990}{\emph{Science} {\bf 357}
  (2017) 1123--1126},
  [\href{http://arxiv.org/abs/https://www.science.org/doi/pdf/10.1126/science.aao0990}{{\tt
  https://www.science.org/doi/pdf/10.1126/science.aao0990}}].

\bibitem{PhysRevD.95.115028}
I.~M. Shoemaker, \emph{Coherent search strategy for beyond standard model
  neutrino interactions},
  \href{http://dx.doi.org/10.1103/PhysRevD.95.115028}{\emph{Phys. Rev. D} {\bf
  95} (Jun, 2017) 115028}.

\bibitem{LIAO201754}
J.~Liao and D.~Marfatia, \emph{Coherent constraints on nonstandard neutrino
  interactions},
  \href{http://dx.doi.org/https://doi.org/10.1016/j.physletb.2017.10.046}{\emph{Physics
  Letters B} {\bf 775} (2017) 54--57}.

\bibitem{PhysRevD.95.055006}
T.~Araki, S.~Hoshino, T.~Ota, J.~Sato and T.~Shimomura, \emph{Detecting the
  ${L}_{\ensuremath{\mu}}\ensuremath{-}{L}_{\ensuremath{\tau}}$ gauge boson at
  belle ii}, \href{http://dx.doi.org/10.1103/PhysRevD.95.055006}{\emph{Phys.
  Rev. D} {\bf 95} (Mar, 2017) 055006}.

\bibitem{PhysRevD.98.015005}
M.~Abdullah, J.~B. Dent, B.~Dutta, G.~L. Kane, S.~Liao and L.~E. Strigari,
  \emph{Coherent elastic neutrino nucleus scattering as a probe of a
  $z\ensuremath{'}$ through kinetic and mass mixing effects},
  \href{http://dx.doi.org/10.1103/PhysRevD.98.015005}{\emph{Phys. Rev. D} {\bf
  98} (Jul, 2018) 015005}.

\bibitem{PhysRevLett.112.231806}
{\scshape ATLAS Collaboration} collaboration, G.~Aad, B.~Abbott, J.~Abdallah,
  S.~Abdel~Khalek, O.~Abdinov, R.~Aben et~al., \emph{Measurements of
  four-lepton production at the $z$ resonance in $pp$ collisions at
  $\sqrt{s}=7$ and 8 tev with atlas},
  \href{http://dx.doi.org/10.1103/PhysRevLett.112.231806}{\emph{Phys. Rev.
  Lett.} {\bf 112} (Jun, 2014) 231806}.

\bibitem{CMS:2012bw}
{\scshape CMS} collaboration, S.~Chatrchyan et~al., \emph{{Observation of Z
  Decays to Four Leptons with the CMS Detector at the LHC}},
  \href{http://dx.doi.org/10.1007/JHEP12(2012)034}{\emph{JHEP} {\bf 12} (2012)
  034}, [\href{http://arxiv.org/abs/1210.3844}{{\tt 1210.3844}}].

\bibitem{ATLAS:2016neq}
{\scshape ATLAS, CMS} collaboration, G.~Aad et~al., \emph{{Measurements of the
  Higgs boson production and decay rates and constraints on its couplings from
  a combined ATLAS and CMS analysis of the LHC pp collision data at $
  \sqrt{s}=7 $ and 8 TeV}},
  \href{http://dx.doi.org/10.1007/JHEP08(2016)045}{\emph{JHEP} {\bf 08} (2016)
  045}, [\href{http://arxiv.org/abs/1606.02266}{{\tt 1606.02266}}].

\bibitem{ATLAS:2019cid}
{\scshape ATLAS} collaboration, M.~Aaboud et~al., \emph{{Combination of
  searches for invisible Higgs boson decays with the ATLAS experiment}},
  \href{http://dx.doi.org/10.1103/PhysRevLett.122.231801}{\emph{Phys. Rev.
  Lett.} {\bf 122} (2019) 231801}, [\href{http://arxiv.org/abs/1904.05105}{{\tt
  1904.05105}}].

\bibitem{Gninenko:2001hx}
S.~N. Gninenko and N.~V. Krasnikov, \emph{{The Muon anomalous magnetic moment
  and a new light gauge boson}},
  \href{http://dx.doi.org/10.1016/S0370-2693(01)00693-1}{\emph{Phys. Lett. B}
  {\bf 513} (2001) 119}, [\href{http://arxiv.org/abs/hep-ph/0102222}{{\tt
  hep-ph/0102222}}].

\bibitem{Baek:2001kca}
S.~Baek, N.~G. Deshpande, X.~G. He and P.~Ko, \emph{{Muon anomalous g-2 and
  gauged L(muon) - L(tau) models}},
  \href{http://dx.doi.org/10.1103/PhysRevD.64.055006}{\emph{Phys. Rev. D} {\bf
  64} (2001) 055006}, [\href{http://arxiv.org/abs/hep-ph/0104141}{{\tt
  hep-ph/0104141}}].

\bibitem{Muong-2:2021ojo}
{\scshape Muon g-2} collaboration, B.~Abi et~al., \emph{{Measurement of the
  Positive Muon Anomalous Magnetic Moment to 0.46 ppm}},
  \href{http://dx.doi.org/10.1103/PhysRevLett.126.141801}{\emph{Phys. Rev.
  Lett.} {\bf 126} (2021) 141801}, [\href{http://arxiv.org/abs/2104.03281}{{\tt
  2104.03281}}].

\bibitem{Minkowski:1977sc}
P.~Minkowski, \emph{{$\mu \to e\gamma$ at a Rate of One Out of $10^{9}$ Muon
  Decays?}}, \href{http://dx.doi.org/10.1016/0370-2693(77)90435-X}{\emph{Phys.
  Lett. B} {\bf 67} (1977) 421--428}.

\bibitem{GellMann:1980vs}
M.~Gell-Mann, P.~Ramond and R.~Slansky, \emph{{Complex Spinors and Unified
  Theories}}, {\emph{Conf. Proc. C} {\bf 790927} (1979) 315--321},
  [\href{http://arxiv.org/abs/1306.4669}{{\tt 1306.4669}}].

\bibitem{Mohapatra:1979ia}
R.~N. Mohapatra and G.~Senjanovic, \emph{{Neutrino Mass and Spontaneous Parity
  Nonconservation}},
  \href{http://dx.doi.org/10.1103/PhysRevLett.44.912}{\emph{Phys. Rev. Lett.}
  {\bf 44} (1980) 912}.

\bibitem{PhysRevD.22.2227}
J.~Schechter and J.~W.~F. Valle, \emph{Neutrino masses in su(2)
  \ensuremath{\bigotimes} u(1) theories},
  \href{http://dx.doi.org/10.1103/PhysRevD.22.2227}{\emph{Phys. Rev. D} {\bf
  22} (Nov, 1980) 2227--2235}.

\bibitem{Schechter:1981cv}
J.~Schechter and J.~W.~F. Valle, \emph{{Neutrino Decay and Spontaneous
  Violation of Lepton Number}},
  \href{http://dx.doi.org/10.1103/PhysRevD.25.774}{\emph{Phys. Rev. D} {\bf 25}
  (1982) 774}.

\end{thebibliography}\endgroup

\end{document}